\documentclass[twocolumn,aps,prb]{revtex4}
\usepackage{graphics,graphicx}
\usepackage{dcolumn}
\usepackage{bm}

\newcommand{\Ref}[1]{Ref.~\onlinecite{#1}}

\def\eb{\begin{equation}}   
\def\ee{\end{equation}}     
\def\ea#1{\begin{eqnarray} #1 \end{eqnarray}}   

\def\shro{Schr\"odinger}

\def\of#1{\left(#1\right)}

\def\prtsq#1{{\partial^2 \over \partial {#1}^2}}

\def\H{\hat{H}}

\def\ra{\rightarrow}

\def\eq#1{Eq.~(\ref{#1})}
\def\eqs#1#2{Eqs.~(\ref{#1}) and (\ref{#2})}

\def\sof#1{\left[ {#1} \right]}
\def\bof#1{\left\{ {#1} \right\}}

\def\Dlt{\Delta}
\def\H{\hat{H}}

\def\Pp{\Psi_+}
\def\Pm{\Psi_-}
\def\Ppm{\Psi_\pm}

\def\pp{\psi_+}
\def\pem{\psi_-}
\def\ppm{\psi_\pm}

\def\pLpm{\psi_{L\pm}}
\def\pRpm{\psi_{R\pm}}
\def\Fp{\Phi_+^E}
\def\Fm{\Phi_-^E}
\def\Fpm{\Phi_\pm^E}

\def\fp{\phi_+^E}
\def\f{\phi^E}
\def\fm{\phi_-^E}
\def\fpm{\phi_\pm^E}
\def\fmp{\phi_\mp^E}
\def\ord#1{{\cal O}(#1)}

\def\tmax{t_{\text{max}}}
\def\Veff{V^{\text{eff}}}
\def\Emin{E_{\text{min}}}
\def\pmin{p_{\text{min}}}
\def\tmax{t_{\text{max}}}

\begin{document}

\title{Reconciling Semiclassical and Bohmian Mechanics: \\
V. Wavepacket Dynamics}

\author{Bill Poirier}
\affiliation{Department of Chemistry and Biochemistry, and
         Department of Physics, \\
          Texas Tech University, Box 41061,
         Lubbock, Texas 79409-1061}
\email{Bill.Poirier@ttu.edu}

\begin{abstract}

In previous articles [J. Chem. Phys. {\bf 121} 4501 (2004), J. Chem.
Phys. {\bf 124} 034115 (2006), J. Chem. Phys. {\bf 124} 034116
(2006), J. Phys. Chem. A {\bf 111} 10400 (2007)] a bipolar
counter-propagating wave decomposition, $\Psi = \Psi_+ + \Psi_-$,
was presented for stationary states $\Psi$ of the one-dimensional
\shro\ equation, such that the components $\Ppm$ approach their
semiclassical WKB analogs in the large action limit. The
corresponding bipolar quantum trajectories are classical-like and
well-behaved, even when $\Psi$ has many nodes, or is wildly
oscillatory. In this paper, the method is generalized for
time-dependent wavepacket dynamics applications, and applied to
several benchmark problems, including multisurface systems with
nonadiabatic coupling.

\end{abstract}

\maketitle



\section{INTRODUCTION}
\label{intro}

Trajectory interpretations of quantum mechanics have existed since
the beginning of the quantum theory---even predating the \shro\ equation
itself. Indeed, one such approach survives today in the form of the
Jeffrey-Wentzel-Kramers-Brillouin (JWKB) approximation, or more generally,
semiclassical mechanics.\cite{tannor,froman,berry72}
In this approach, a time-evolving
quantum wavepacket is treated as a statistical ensemble of classical
trajectories that ``carry'' approximate quantum information, i.e. complex
amplitudes. In the early 1950s, hearkening back to the earlier pioneers,
D. Bohm and coworkers developed a conceptually similar trajectory
interpretation of the {\em exact} quantum
theory,\cite{madelung26,bohm52a,bohm52b} introducing the so-called
``quantum potential'' to guide the resultant quantum trajectory dynamics.
In the intervening years, quantum trajectory methods (QTM)s have been
used ``analytically''---to provide insight into solved time-dependent
quantum wavepacket propagation
problems\cite{holland,wu93,sanz02,wangz01}---and more recently, as a
``synthetic'' tool---to actually solve the time-dependent \shro\
equation (TDSE) itself.\cite{wyatt,lopreore99,mayor99,shalashilin00,wyatt01,wyatt01b,wyatt02b,bittner02b,hughes03}
Note that in this paper, ``QTM'' refers to the original quantum trajectory
method based on standard Bohmian mechanics, as developed by Wyatt and
coworkers,\cite{wyatt} {\em as well as} the various offshoots and approximations
that have developed in the intervening years.

The standard Bohmian formulation uses an amplitude-phase
decomposition of the wavefunction, $\psi$, which in one dimension (1D)
takes the form
\eb
     \psi(x,t) = R(x,t) e^{i S(x,t)/\hbar}. \label{oneLM}
\ee
This representation has been called ``unipolar,''\cite{wyatt} because the field
functions, $R(x,t)$ and $S(x,t)$, are single-valued at all positions, $x$,
and times, $t$. The field functions, and the resultant quantum trajectories,
are generally smooth and classical-like---provided the true potential,
$V(x)$, is slowly-varying, {\em and} $\psi(x,t)$ exhibits no interference.
However, interference introduces non-classical-like oscillations in
$R(x,t)$ and $S(x,t)$, which in turn lead to severe numerical difficulties
for QTM calculations---collectively referred to as
``the node problem.''\cite{wyatt,wyatt01b} Despite substantial
progress,\cite{babyuk04,trahan03,kendrick03,pauler04} the node problem continues
to be the most formidable roadblock impeding the progress of QTM's as a
general and robust tool for exact quantum scattering applications.

As a promising remedy to the node problem, this paper is the fifth in a
series\cite{poirier04bohmI,poirier06bohmII,poirier06bohmIII,poirier07bohmIV}
exploring the use of {\em bipolar} decompositions of the wavefunction, i.e.
\eb
     \psi = \pp +\pem,  \label{psitot}
\ee
such that the quantum trajectories associated with the bipolar
wavefunction components, $\ppm(x,t)$, are well-behaved and classical-like,
even when the unipolar quantum trajectories associated with $\psi(x,t)$
itself are not---e.g., when $\psi(x,t)$ exhibits interference. In fact,
by generalizing Bohmian mechanics for multipolar decompositions such as
\eq{psitot} above, it becomes possible to achieve
{\em classical correspondence}---i.e., trajectories and field functions
that approach their classical counterparts in the classical limit of large
mass, energy, and/or action---which is not in general possible for any
unipolar Bohmian treatment.

Most QTM papers found in the literature concern themselves with
time-dependent localized wavepacket dynamics, as indeed, is also
true of this fifth paper in the bipolar series. However, this
paper represents a stark departure from the previous
four, all of which pertain to stationary solutions of the the time-{\em
independent} \shro\ equation---albeit obtained in a
pseudo-time-dependent manner using delocalized counter-propagating wave
components. For 1D stationary states, the two $\ppm(x,t)$ components
correspond to approximate semiclassical analogs,\cite{froman,berry72}
of which there are always two at every classically-allowed point in
space and time---thus justifying a bipolar QTM treatment in this context.
If there are classical turning points demarcating classically
allowed and forbidden regions, these are known {\em a priori}, and
do not change over time.

In contrast, the semiclassical treatment of localized {\em wavepacket} dynamics,
even in 1D, is considerably more complicated than for stationary
states---rendering it far more difficult to arrive at a suitable QTM analog.
In particular, the time-evolving semiclassical field functions can become
multivalued over regions of $x$ and $t$ in a highly non-trivial way,
due to caustics that form, move, and then disappear over time.\cite{tannor}
The appropriate number of semiclassical components can become one, two, or
even three or more, and in any case, varies nontrivially over $x$ and $t$.
The formation of caustics, in turn, is due to the crossing of neighboring
classical trajectories belonging to a semiclassical wavepacket ensemble.
In stark contrast are quantum trajectories, which, for a single component,
may {\em never} cross---implying single-valuedness for all $x$ and $t$,
and evidently greatly complicating the task of achieving classical
correspondence.

Let us imagine, for instance, that at the initial time, $t_0$,
$\psi(x,t_0)$ is taken to be of the unipolar \eq{oneLM} form.
Though initially single-valued, $\psi(x,t)$ may, at later $t$,
become multivalued over certain regions of $x$, under a semiclassical
treatment. However, since an ensemble of {\em quantum} trajectories does
not develop caustics, how then is a QTM to become similarly
multivalued? This fundamental difficulty persists even under the
usual allowance that quantum and semiclassical wavefunctions are
equivalent only to $\ord{\hbar}$. In fact, it raises the following, purely
semiclassical conundrum, which to the author's best knowledge, has not been
previously addressed: if $\psi(x,t)$ can become multivalued at {\em
later} $t$, why should it never be considered so at $t=t_0$? One practical
reason is that the $\ord{\hbar}$ uniqueness of the semiclassical
representation would be compromised, making it unclear how to proceed.
This answer is not satisfying in any formal sense; yet indeed,
semiclassical theory treats the same Gaussian wavefunction
as {\em single-valued} if interpreted in a wavepacket context,
or {\em double-valued} if regarded as a harmonic oscillator ground state.

With regard to QTM wavepacket dynamics, the multivalued problem
described above can in principle be addressed in a variety of ways.
One simple strategy would be to actually use classical trajectories
for the dynamics---not approximately, as in semiclassical theories,
but rather, with {\em exact} propagation of the (complex-valued) quantum
amplitudes, $R(x,t)$.  This approach can be regarded as an arbitrary
Lagrangian-Eulerian (ALE) method,\cite{wyatt,trahan03,kendrick03}
modified to allow for the formation of caustics and multivalued
fields. Note that probability is no longer conserved along
trajectories---and indeed, must approach zero at the caustics, in
order to avoid infinite probability density. In practice, this
condition itself causes numerical instabilities, and in any case
would be unfeasible to implement for large
systems.\cite{poirier07bohmVunpub1}

Alternatively---and seemingly contrary to semiclassical
behavior---one might opt to adhere strictly to a globally bipolar
decomposition of the \eq{psitot} form, in order to facilitate
comparison with the bipolar stationary state theories of
papers I--IV in the
series.\cite{poirier04bohmI,poirier06bohmII,poirier06bohmIII,poirier07bohmIV}
Even in this relatively restricted
context, however, the ``correct'' bipolar wavepacket generalization
is not necessarily obvious.
In the previous work, for instance, both of the $\pp$ and $\pem$ field
functions are symmetric for stationary bound states,\cite{poirier04bohmI} whereas
only $S$ is symmetric [i.e., $S_-(x) = -S_+(x)$] for the more general case of
stationary scattering states.\cite{poirier06bohmII,poirier06bohmIII}
For bipolar wavepacket dynamics, {\em neither} field function provides
us with simplifying symmetry; moreover, we evidently have no direct recourse to
semiclassical mechanics, which was previously relied upon as a guide.

One way to state the problem is as follows: a complete specification
of the bipolar wavepacket dynamics generally requires {\em four}
independent real-valued time-evolution equations. Two of these are
automatically provided by the TDSE, but how are the remaining two to
be chosen? One promising avenue, which we have explored
considerably,\cite{poirier07bohmVunpub2} is to adopt the combined flux
continuity condition\cite{poirier06bohmIII,poirier07bohmIV,poirier07bohmalg}
[\eq{fluxrel}] as the third equation. In fact, if {\em imaginary}
flux is considered,\cite{poirier07bohmcomplex} the fourth and final equation
can also be obtained. Unfortunately, these equations do not provide
satisfactory results, in that over time, the individual $\ppm(x,t)$
components themselves develop interference oscillations and
nodes---thus defeating the purpose of a bipolar expansion. More
flexibility can be obtained by dropping the imaginary flux
continuity condition, but to date, all such efforts have also been
unsuccessful\cite{poirier07bohmVunpub2}---either due to
$\ppm(x,t)$ interference, or other equally unsatisfactory behaviors
(Sec.~\ref{additional}).

Ultimately, the most successful bipolar wavepacket generalization scheme
we have considered has also proven to be one of the most conceptually
straightforward---i.e., to expand $\psi(x,t)$ as a superposition of
stationary state solutions, whose
$+$ and $-$ components are then used to determine the wavepacket
$\ppm(x,t)$ via linear superposition.
The idea is simple, but the theoretical development is
somewhat involved. In any event, this approach, which will serve as the
focus of this paper, leads to a node-free, and otherwise remarkably
well-behaved \eq{psitot} decomposition---and moreover, turns out to satisfy
classical correspondence after all. It also leads to time-evolution equations
that are practicable for numerical implementation. Note that bipolar
quantum trajectory formulations for the TDSE have been considered
previously by other authors,\cite{bertoldi00} albeit not in a manner
designed to solve the interference/node problem.

The remainder of this paper is organized as follows. Sec.~\ref{theory}
first presents the requisite background of the bipolar
theory for stationary scattering states (Sec.~\ref{background}),
followed by a derivation of the new wavepacket time-evolution
equations for asymptotically symmetric potentials (Sec.~\ref{evolution}).
Additional properties of the resultant $\ppm(x,t)$ component
wavepacket dynamics are described in Sec.~\ref{additional}.
Generalizations for asymptotically asymmetric and multisurface
applications are then provided in Secs.~\ref{asymmetric}
and~\ref{multisurf}, respectively. Results and discussion, for
benchmark applications of each type described above, are then
presented in Sec.~\ref{results}. Finally, a summary and concluding
remarks are given in Sec.~\ref{conclusion}.


\section{THEORY}
\label{theory}

\subsection{Background}
\label{background}

Consider the two bipolar components, $\fpm(x)$, associated with a 1D
stationary scattering state solution of the \shro\ equation,
$\f(x)= \fp(x) + \fm(x)$,
with energy, $E$, and left-incident boundary conditions. The solution
components, $\fpm(x)$, are exact quantum analogs of a type of
semiclassical JWKB approximation resulting from the ``generalized
Fr\"oman''
approach.\cite{froman,poirier06bohmIII,poirier07bohmIV,poirier07bohmalg}
Note that for the remainder of this paper, $\phi$ is used in the
context of stationary state wavefunctions, whereas $\psi$ is
reserved for localized wavepacket dynamics. The solution components
$\fpm(x)$ behave as (right/left) traveling plane waves in both
asymptotes. As $x$ sweeps through the potential interaction region,
the solution component $\fp(x)$ varies smoothly from incident to transmitted
wave, whereas $\fm(x)$ varies smoothly from reflected wave
to zero.

In several previous
articles,\cite{poirier06bohmII,poirier06bohmIII,poirier07bohmIV,poirier07bohmalg}
an extremely accurate, efficient, and robust 1D numerical algorithm was
developed for computing $\fpm(x)$, and thus $\f(x)$. The algorithm is a
time-dependent relaxation method, for which the initial $\f = \fp$ is a
plane wave. Over time, a reflected wave, $\fm(x,t)$,
comes into being through interaction region coupling due to the
potential energy, and eventually, $\f(x,t)$ relaxes to the true
stationary scattering solution.

From \Ref{poirier06bohmIII} Eq.~(12), the $\fpm(x,t)$ time-evolution equations
are
\eb
     {\partial \fpm \over \partial t} = \mp {p \over m} {\fpm}' +
                         {i \over \hbar} \of{E-V}\fpm -
                         {i \over \hbar} V \fmp \label{fdot},
\ee
where primes denote spatial differentiation, $m$ is the mass,
$p=\sqrt{2mE}$ is the magnitude of the asymptotic momentum, and
asymptotically symmetric potentials are presumed [$V(x) \ra 0$ as
$x\ra\pm\infty$]. The initial value conditions are given by
\ea{
     {\fp}^0(x) & = &\fp(x,t_0) = \exp\!\sof{ {i p x\over\hbar} -
                                             {i E t_0 \over \hbar}} \\
     {\fm}^0(x) & = &\fm(x,t_0) =0, \nonumber}
where $t=t_0\ra-\infty$ is the initial time.
At the left and right coordinate limits, i.e. $x= x_L \ra
-\infty$ and $x=x_R \ra +\infty$, respectively, the boundary
conditions are:
\ea{
     \fp(x_L,t) & = & \exp\!\sof{ {i p x_L\over\hbar} - {i
           E t\over \hbar}} \label{bc} \\
     \fm(x_R,t) & = & 0 \nonumber
}

In general, \eq{fdot} does {\em not} satisfy the TDSE, in
that $\of{{\partial \f / \partial t}} \ne -(i/\hbar) \H \f$ for
all times $t$, where $\H$ is the usual \shro\
equation Hamiltonian. This is consistent with the interpretation of
this approach as a ``revelatory'' or ``relaxation''
method,\cite{poirier06bohmII,poirier06bohmIII,poirier07bohmalg}
but implies that \eq{fdot} itself
cannot be used as a basis for deriving wavepacket time-evolution
equations, for which the TDSE must be satisfied at all $t$.

\subsection{Wavepacket time-evolution equations}
\label{evolution}

We therefore consider the asymptotically large time limit, $t\ra
+\infty$, in which \eq{fdot} not only satisfies the TDSE, but
relaxes to the exact stationary solution, so that
\eb
     {\partial \f \over \partial t} = -{i \over \hbar} E \f \quad
     \text{and} \quad
     {\partial \fpm \over \partial t} = -{i \over \hbar} E \fpm
     \quad \text{as $t\ra +\infty$}. \label{tasymp}
\ee
Substituting \eq{tasymp} into \eq{fdot} and rearranging yields
the following time-independent expressions for the spatial
derivatives of the solution ($t\ra+\infty$) $\fpm(x)$:
\eb
     {\fpm}' = \pm {i \over \hbar} \,p\, \fpm
             \mp  {i \over \hbar} \of{{m \over p}} V \of{\fp + \fm}.
      \label{fslash}
\ee

Equation~(\ref{fslash}) above is consistent with
\Ref{poirier07bohmalg} Eq.~(12), with $v = p/m = \sqrt{2E/m}$.
Differentiating \eq{fslash} with respect to $x$, and then using
\eq{fslash} to substitute for ${\fpm}'$ in the resulting right hand
side yields: \eb
     {\fpm}'' = - {2 m \over \hbar^2}\of{E-V}\fpm
        \mp  {i \over \hbar} \of{{m \over p}} V' \of{\fp + \fm}. \label{fslash2}
\ee
Equation~\ref{fslash2} can then be used to obtain
\eb
    \H \fpm = E \fpm \pm {i \hbar \over 2 p} V' \of{\fp + \fm}.
\ee
Substituting \eq{tasymp} for  $E\fpm$ above then results in the following, new
time-evolution equations:
\eb
     {\partial \fpm \over \partial t} =
                        - {i \over \hbar} \H \fpm \mp
                         {V' \over 2p} \of{\fp + \fm} \label{fdot2}
\ee

A subtle shift has occured in the transformation from \eq{fdot} to
\eq{fdot2}. First, the latter manifestly satisfies the TDSE at
{\em all} times $t$. Second, what constitutes
the coupling contribution (i.e. the last term) in \eq{fdot} is
{\em not} equivalent to that of \eq{fdot2}. Most strikingly, the latter
coupling is proportional to $V'$ rather than to $V$ itself. This implies that
the coupling vanishes in both of the coordinate asymptotic limits,
$x_L$ and $x_R$, {\em even} when $V(x)$ is not taken to be asymptotically
symmetric. This represents quite an improvement over \eq{fdot}, which
cannot be applied to asymmetric potentials because there is asymptotic
coupling in at least one coordinate limit.\cite{poirier06bohmIII,poirier07bohmalg}
Although \eq{fdot2} can be used with asymptotically asymmetric potentials,
there are nontrivial ramifications for wavepacket dynamics
(Sec.~\ref{asymmetric}).
The final, and perhaps most important, observation that will be made
regarding the coupling contribution to \eq{fdot2} is that it is {\em
directly proportional to the JWKB quantity denoting the classical
limit}---i.e., $|\lambda V'|$, where $\lambda = 2\pi \hbar / p$ is
the de Broglie wavelength.\cite{liboff} In particular, the classical limit
is obtained when $|\lambda V'|\ll p^2 / 2m$---i.e., $(V'/p)\ra 0$.
According to \eq{fdot2}, the coupling {\em vanishes} in the classical
limit, in which the $\fpm$ themselves approach \shro\ equation solutions.
In this manner, classical correspondence is established for the new
time-evolution equations.

All of the above still refers to delocalized left-incident
stationary scattering states $\f$, at definite energies $E$. For every
positive $E$ value, the methodology uniquely determines a
corresponding $\f$ and $\fpm$. In generalizing for (left-incident)
localized wavepacket dynamics, an eminently sensible strategy is to
decompose the wavepacket $\psi$ as an orthonormal expansion in the
stationary states $\f$---an expansion which, in turn, is applied to the
$\fpm$ components themselves, to uniquely determine $\ppm$ and the
\eq{psitot} bipolar decomposition. Thus,
\ea{
     \psi(x,t) & = & \int_0^\infty a(E) \f(x,t)\, dE
                           \quad \text{and}  \label{psidec} \\
     \ppm(x,t) & = & \int_0^\infty a(E) \fpm(x,t)\, dE .
          \nonumber }
In principle, the above equations enable $\ppm(x,t)$ to be completely
determined---provided the initial wavepacket, $\psi^0(x)= \psi(x,t_0)$,
is specified, and all of the solution $\fpm(x,t)$'s are known {\em a priori}.
In practice, the latter requirement defeats the purpose of
doing localized wavepacket dynamics---i.e.,
to avoid explicit calculation of the delocalized $\f$ states.

Consequently, we apply the \eq{psidec} expansion to both sides
of \eq{fdot2}, in order to directly derive time-evolution equations
for $\ppm(x,t)$. We require that the explicit integrations over $dE$
be tractable, so that $\f$ and $\fpm$ not appear explicitly in
the final results for $\partial \ppm / \partial t$. This in turn requires
that the right-hand-side of \eq{fdot2} exhibit no explicit dependence
on $E$ or $p$, which---due to the coupling term---is seen not to be
satisfied. To make progress, we use the identity,
\eb
     {\f}' = {i p \over  \hbar} \of{\f_{+} - \f_{-}}, \label{fprime}
\ee
obtained from \Ref{poirier06bohmIII} Eq.~(9), or from
\Ref{poirier07bohmalg} Eq.~(11), or by using \eq{fslash} to add
${\fp}'+{\fm}'$. Substituting the integral of \eq{fprime} with respect to
$x$ into \eq{fdot2} then yields
\ea{
     {\partial \fpm \over \partial t} & = &
              - {i \over \hbar} \sof{\H \fpm \pm
              {V' \over 2} \of{\Fp - \Fm}},\label{fdot3} \\
              \text{where} \quad \Fpm(x) & = & \int_{-\infty}^x
              \fpm(x')\,dx', \nonumber}
apart from a term proportional to $\fpm(x\ra -\infty)$ that vanishes
when integrated over $E$ via \eq{psidec} (Dirichlet wavepacket
boundary conditions, Sec.~\ref{results}). Since  \eq{fdot3} above
has no explicit dependence on $E$ or $p$, applying the \eq{psidec}
expansion to \eq{fdot3} leads at once to the following wavepacket
time-evolution equations:
\ea{
     {\partial \ppm \over \partial t} & = &
              - {i \over \hbar} \sof{\H \ppm \pm
              {V' \over 2} \of{\Pp - \Pm}},\label{pdot} \\
              \text{where} \quad \Ppm(x) & = & \int_{-\infty}^x
              \ppm(x')\,dx' \label{pint}}

Equation~(\ref{pdot}) can be directly integrated over time, to
determine the dynamics of the bipolar wavepacket components,
$\ppm(x,t)$. In addition to the usual TDSE Hamiltonian contribution,
\eq{pdot} includes a coupling contribution that is proportional to
$V'$ (and independent of mass). As in the case of \eq{fdot3},
this implies that the asymptotic $x\ra \pm\infty$ coupling vanishes,
even for asymmetric potentials. Since $V(x)$ also vanishes
asymptotically, we thus find that $\ppm(x\ra\pm\infty,t)$ evolves
under free-particle propagation. Note that throughout the $x$ coordinate
range, the $\ppm$ coupling terms are equal and opposite, so that
$\of{{\partial \psi /\partial t}} = -(i/\hbar) \H \psi$. Thus, \eq{pdot}
satisfies the TDSE at all $t$.

\subsection{Additional properties}
\label{additional}

Having defined a set of wavepacket time-evolution equations [\eq{pdot}],
we next consider whether these give rise to well-behaved $\ppm(x,t)$
components at all $x$ and $t$. In general, a great range
of behaviors are possible for nonstationary state dynamics in 1D, many
of which are undesirable. We therefore first stipulate that---apart from exhibiting
interference---$\psi(x,t)$ itself be well-behaved. By this we mean that
$\psi(x,t)$ is normalized to unity and well-localized at all times $t$,
consisting of a single left-incident wavepacket at the initial time $t=t_0$,
and of well-separated left- and right-moving reflected and transmitted branches,
respectively, at the final time $t= t_f\ra+\infty$. Under these assumptions
for the $\psi(x,t)$ wavepacket dynamics, we define well-behaved
{\em components}, $\ppm(x,t)$, as those that satisfy the following three conditions:
\begin{itemize}
\item{{\em Condition 1:} Perfect asymptotic separation at $t_0$ and $t_f$.}
\item{{\em Condition 2:} Well-localized $\ppm(x,t)$ and $\Ppm(x,t)$
at all $t$.}
\item{{\em Condition 3:} Node-free components, $\ppm(x,t)$, at all $t$.}
\end{itemize}
Condition 1. means that the initial left-incident wavepacket
consists solely of $\pp$---i.e.,  $ \pp^0(x) = \psi^0(x) = \psi(x,t_0)$,
and $\pem^0(x) = 0$. It also means that at the asymptotically large
final time $t_f$, $\pp^f(x)= \pp(x,t_f)$ becomes the right-moving transmitted
branch of $\psi^f(x) = \psi(x,t_f)$, and $\pem^f(x)= \pem(x,t_f)$ becomes
the left-moving reflected branch. Condition 2. is straightforward,
and absolutely essential, e.g., for multidimensional generalizations.
Condition 3. is expected to hold at
{\em all} $t$---particularly intermediate times, where $\psi(x,t)$
itself may exhibit substantial interference and/or nodes.

For the remainder of this subsection, asymptotically symmetric potentials
are presumed, as defined in Sec.~\ref{background}. At the initial time $t_0$,
the incident wavepacket is localized far to the left of the
potential interaction region, so that $V(x)$ is effectively zero, and the
\eq{bc} plane wave boundary condition accurately describes $\fp(x,t)$---over
the whole asymptotic region where $|\psi^0(x)|^2$ is significant. Thus,
the initial \eq{psidec} stationary state expansion is essentially
{\em identical to a Fourier expansion}---or equivalently, the momentum-space
representation, $\tilde{\psi}^0(p)$. Through the identification $p=\sqrt{2 m E}$,
we find that only the $p>0$ states contribute in the \eq{psidec} expansion
for $\pp^0(x)$; the $p\le0$ states give rise to a left-moving contribution,
which is presumed to be zero at $t=t_0$. Condition 1. thus requires that the
negative momentum contribution to $\psi^0(x)$ be vanishingly small, i.e.
\eb
    \int_{-\infty}^0 |\tilde{\psi}^0(p)|^2 dp \ra 0. \label{pcon}
\ee
This is a very reasonable requirement to impose on $\psi^0(x)$, for it
implies that the initial wavepacket is completely incident upon the
scattering potential center.

At the final time $t_f$, the $a(E)$ expansion coefficients from \eq{psidec}
are identical to their initial values at $t_0$. Moreover, the reflected and
transmitted wavepacket branches are localized far to the left and right,
respectively, of the interaction region, so that once again, the $\f(x,t)$
are effectively plane waves.\cite{poirier06bohmIII,poirier07bohmalg}
However, these asymptotic plane waves are no longer characterized by the
standard unit normalization of \eq{bc}, but must instead be weighted by
the $E$-dependent reflection and transmission amplitudes, $R(E)$ and $T(E)$,
when determining the Fourier components, $\tilde{\psi}^f(p)$. Note that
since $\fm(x_R,t) = 0$ [\eq{bc}], there can be no $\pem$ component
in the right asymptote, implying that the right-moving transmitted
branch at $t_f$  must consist only of a $\pp^f(x)$
contribution. Similar arguments can be used to demonstrate that the
left-moving reflected branch at $t_f$ must consist only of $\pem^f(x)$.
We thus find that Condition 1. above is formally satisfied at both $t_0$
and $t_f$.

Regarding Condition 2., here again we make use of \eq{pcon}. If \eq{pcon}
were {\em not} satisfied, then $\tilde{\psi}(p=0,t)$ would in general be
nonzero. Since $\Psi(x_R,t) = \sqrt{2 \pi \hbar} \tilde{\psi}(p=0,t)$
[from \eq{pint}], the right-asymptotic value of $\Psi(x_R,t)$ would approach
a constant (in $x$), nonzero value. Thus, $\Psi(x,t)$ would be
{\em delocalized}---even if $\psi(x,t)$ itself were localized, as it is
initially. Over time, moreover, due to the coupling term in \eq{pdot},
the $\ppm(x,t)$ would themselves eventually become delocalized,
even if $\psi(x,t)$ itself were not---clearly, an untenable situation, in
violation of Condition 2. Conversely to the above scenario, the fact that
\eq{pcon} is true implies that $\Ppm(x,t)$, and therefore $\ppm(x,t)$, are
localized---not {\em only} at asymptotic times, but at {\em all} times.
Thus, Condition 2. is also formally satisfied.

\begin{figure}
\includegraphics[scale=0.95]{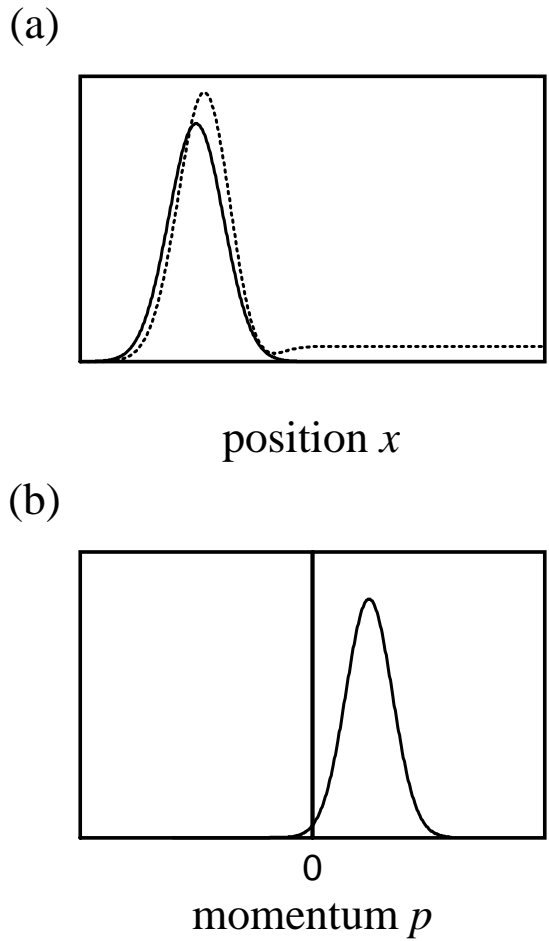}
        \caption{Schematic indicating properties of initial wavepacket, $\psi^0$:
                 (a) position space; (b) momentum space. The solid line in (a) represents
                 $\rho^0(x)= |\psi^0(x)|^2$, the initial wavepacket density, taken to
                 be a Gaussian centered far to the left of the interaction region. The
                 dotted line in (a) represents $|\Psi^0(x)|^2$, where
                 $\Psi(x)  =  \int_{-\infty}^x \psi(x')\,dx'$, as per \eq{pint}. Note that
                 $\Psi(x)$ is {\em not} localized, owing to the small constant-valued tail
                 that extends towards $x=+\infty$. This is because the Fourier transform,
                 $\tilde{\psi}^0(p)$, does not satisfy \eq{pcon} perfectly---as seen in (b),
                 a plot of $|\tilde{\psi}^0(p)|^2$.}
        \label{tailfig}
\end{figure}

In practice, \eq{pcon} is never perfectly satisfied (even in the $t_0\ra-\infty$
limit), but is only approximately correct, to some desired level of numerical
accuracy. The true $\Ppm(x,t)$ will have a small delocalized
constant-valued tail---extending to the right towards $x\ra\infty$
for the definite integration convention of \eq{pint}, as indicated in
Fig.~\ref{tailfig}, but non-vanishing no matter which integration limits are
adopted. For reasonable ${\tilde\psi}^0(p)$ distributions however, the magnitude
of these tails can easily be made arbitrarily small (and therefore insignificant)
simply by shifting to ${\tilde\psi}^0(p-p_0)$ for sufficiently large $p_0$.

Finally, we address the all-important Condition 3. Though we can
offer no formal proof at present that this condition is always
satisfied, it must certainly be true in the classical limit, in
which $(V'/p)$ approaches zero. More generally, if
Condition 3. is satisfied for the individual $\fpm$ components---as
has been demonstrated for a great range and variety of test
applications\cite{poirier06bohmII,poirier06bohmIII,poirier07bohmIV,poirier07bohmalg}---then
it is reasonable to expect this property to also be preserved under
the \eq{psidec} expansion. In any event, Condition 3. is verified for
each of the test cases considered in Sec.~\ref{results}.

On the other hand, one nice property of the $\fpm$ under \eq{fdot}
that is definitely {\em not} extended to the $\ppm$ under \eq{pdot}
is that of combined flux continuity.
In other words, unlike all previous bipolar formulations for stationary
state dynamics,\cite{poirier06bohmIII,poirier07bohmIV,poirier07bohmalg}
we find here that
\eb
    \of{{\partial \rho_+ \over \partial t} +
    {\partial \rho_- \over \partial t}}
    \ne -\of{{j_+}' + {j_-}'},\label{fluxrel}
\ee
where $\rho_\pm = |\ppm|^2$ are component densities, and
$j_\pm = \rho_\pm ({S_\pm}'/m)$ are component fluxes, defined in the
{\em standard} quantum manner via
\eb
    \ppm(x,t) = R_\pm(x,t) \exp\sof{i S_\pm(x,t)/\hbar}.
    \label{psipm}
\ee

The appearance of the standard flux in the above expressions
represents a departure from the stationary state formalism---for
which predetermined classical velocities, $\pm v$, are used rather
than $(S_\pm'/m)$---and a move back towards standard Bohmian
mechanics.\cite{bohm52a,bohm52b,wyatt} On the other hand, the fact that the
\eq{fluxrel} combined flux continuity condition is not satisfied
might lead one to argue that the $(S_\pm'/m)$ velocity field is
inappropriate here, and that some other choice---perhaps some
nontrivial generalization of the stationary state velocities---would
lead to an \eq{fluxrel}-type equality. In fact this is incorrect---for
it can be shown (Sec.~\ref{eckartproton}) that $\int_{-\infty}^{+\infty}
\of{\rho_+ + \rho_-}\, dx$ is not conserved over time, implying the
\eq{fluxrel} inequality {\em regardless} of the particular choice
of velocity field. In any event, the standard Bohmian velocity field
is naturally obtained when \eq{pdot} is used to derive time-evolution
equations for the component densities:
\eb
     {\partial \rho_\pm \over \partial t} = - {j_\pm}' \pm
              {V' \over \hbar} \text{Im} \sof{\ppm^*
              \of{\Pp - \Pm}},\label{rhodot}
\ee
From \eq{rhodot} above, \eq{fluxrel} is easily obtained. Note that
despite the \eq{fluxrel} inequality, $\int_{-\infty}^{+\infty}
\of{\rho_+^0 + \rho_-^0}\, dx = \int_{-\infty}^{+\infty}
\of{\rho_+^f + \rho_-^f}\, dx = \int_{-\infty}^{+\infty} \rho\,dx = 1$,
so that globally over time, the $\of{\rho_+ + \rho_-}$ probability
{\em is} conserved.

\subsection{Asymptotically asymmetric potentials}
\label{asymmetric}

Our next task is to generalize the previous discussion for the case
of asymptotically asymmetric potentials. To be completely general,
we allow the left asymptotic value, $V_L = V(x_L)$, and right asymptotic
value, $V_R = V(x_R)$, to be completely arbitrary---i.e., $V_L \ne V_R$,
and neither $V_L$ nor $V_R$ need be zero. In this context, it is
straightforward to generalize \eq{fdot} for {\em either} of the two
asymptotic potential values, but not both simultaneously. Essentially,
this is done by adopting $\Veff(x)=V^0$ as the effective potential
used to generate classical
trajectories,\cite{poirier06bohmIII,poirier07bohmIV,poirier07bohmalg}
where the constant $V^0$ is chosen to be either $V^0=V_L$ or $V^0=V_R$.
In either case, for the generalized \eq{fdot}, i.e.
\eb
     {\partial \fpm \over \partial t} = \mp {p \over m} {\fpm}' +
                         {i \over \hbar} \of{E-V-V^0}\fpm -
                         {i \over \hbar} (V-V^0) \fmp \label{fdot4}
\ee
[with $p=\sqrt{2m(E-V^0)}$], coupling vanishes in one $x$ asymptote,
but not the other.

As explored in previous papers,\cite{poirier06bohmIII,poirier07bohmalg}
two natural remedies for the asymptotic coupling dilemma are considered:
(1) define a smoothly varying effective potential $\Veff(x)$ such
that $\Veff(x_{L/R}) = V_{L/R}$; (2) define a discontinuous transition
at an intermediate dividing point $x_D$, so that
$\Veff(x) = V_L + (V_R-V_L)\Theta(x-x_D)$, where $\Theta()$ is
the (heaviside) step function. With respect to deriving wavepacket
time-evolution equations as per Sec.~\ref{evolution}, option (1)
poses severe difficulties, in that it is not clear how to recouch
the relevant equations\cite{poirier07bohmalg} to avoid explicit dependence
on $E$ and/or $p$. Option (2) on the other hand, is straightforward,
as we now demonstrate.

The key property is that \eq{fdot4}---whether for $V^0=V_L$, $V^0=V_R$, or
an arbitrary $V^0$ value---leads to {\em exactly} the same
wavepacket evolution equations as for $V^0=0$, i.e., \eq{pdot}.
However, the resultant $\ppm$ components are $V^0$-dependent.
Thus, the $V^0=V_L$ components $\pLpm$, and the $V^0=V_R$
components $\pRpm$, constitute {\em distinct} bipolar decompositions,
each satisfying \eqs{psitot}{pdot}. This can only be true provided
the initial (and final) conditions are different, i.e.
$\pLpm^0 \ne \pRpm^0$ and $\pLpm^f \ne \pRpm^f$,
which in turn implies that Condition 1. (Sec.~\ref{additional})
must be false. In fact, we still
find that $\psi_{L+}^0(x) = \psi^0(x)$ and $\psi_{L-}^0(x) = 0$,
but both $\pRpm^0(x)\ne 0$. Similarly, at $t_f$, the transmitted
branch of $\psi^f(x)$ equals $\psi_{R+}^f(x)$ [no
$\psi_{R-}^f(x)$ contribution], but when expanded instead in terms of
$\pLpm^f(x)$, includes nonzero contributions from both. The
reflected branch of $\psi^f(x)$ equals $\psi_{L-}^f(x)$.

In accord with option (2) above, it is natural to
define a single bipolar decomposition, $\ppm$, that satisfies all
three conditions of Sec.~\ref{additional}, by ``gluing'' together
the asymptotic solutions at the dividing point, $x_D$, as follows:
\eb
    \ppm(x,t) = \Theta(x_D-x) \pLpm(x,t) + \Theta(x-x_D) \pRpm(x,t)
    \label{join}
\ee
Such a procedure is analogous to other dividing surface methods,
commonly used in reactive scattering applications.\cite{seideman92a,rom92}
It is not known at present how to time-evolve \eq{join} directly,
i.e. without recourse to separate calculations for $\pLpm$ and $\pRpm$
over all $x$ and $t$. However, the latter is straightforward to achieve
in practice---requiring, in addition to \eq{pdot}, only the specific
initial conditions $\pRpm^0(x)$, which are derived below.

First, in the \eq{psidec} expansion, the lower limit of the integration
must be replaced with $\Emin =\max\of{V_L,V_R}$, as reactive scattering
does not occur at energies below $\Emin$. For $\psi_{L+}^0(x) = \psi^0(x)$,
this expansion is still equivalent to a Fourier expansion, except that the
minimum allowed (left) momentum value is
\eb
     \pmin = \cases{0 & if $V_R<V_L$; \cr
                    \sqrt{2 m (V_R-V_L)} & otherwise.\cr}
\ee
Thus, the upper limit in \eq{pcon} must be replaced with $\pmin$ rather
than 0, in order that Condition 2. be satisfied. In any event, both
$\tilde{\psi}^0(p)$, and the $a(E)$ expansion coefficients in
\eq{psidec}, can be computed explicitly from $\psi^0(x)$ via straightforward
Fourier transform.

The next step is to relate the $\f_{R\pm}$ decomposition for the
stationary state solution, $\f$, to the corresponding
$\f_{L\pm}$ decomposition. Applying \eq{fslash} to $\f_{R\pm}$, and
rearranging, we obtain
\eb
    \f_{R\pm} = {1\over 2}\sof{\f \mp \of{i\hbar \over p_R}{\f}'},
    \label{frslash}
\ee
where $p_R = \sqrt{2m(E-V_R)}$, For purposes of expanding the initial
wavepacket $\psi^0(x)= \psi_{L+}^0(x)$, we can replace $\f$ in \eq{frslash}
with $\f_{L+}$. Moreover, the $x$ range of interest is restricted to the
left asymptote, where the $\f_{L\pm}$ are plane waves of the \eq{bc} form,
but with $p$ replaced by $p_L = \sqrt{2m(E-V_L)}$. Making these substitutions
in \eq{frslash} leads to
\eb
    \f_{R\pm} = {1\over2}\sof{1\pm p_L/p_R}\f_{L+}, \label{fr}
\ee
from which $\pRpm^0(x)$ can be obtained via straightforward inverse
Fourier transform. In particular, \eqs{psidec}{fr} lead to
\ea{
    \pRpm^0(x) & = & \int_{\Emin}^\infty a(E)
           {1\over2}\sof{1\pm \sqrt{{E-V_L\over E-V_R}}}\\
          & & \times \exp\bof{{i\over\hbar} \sof{\sqrt{2m(E-V_L)} x - E t_0}}\, dE.
           \label{pRinit}\nonumber}
Similar arguments can be used to justify the other initial and
final conditions for $\pLpm$ and $\pRpm$, as discussed earlier in
this subsection.

Using the explicit initial value conditions of \eq{pRinit}, propagation
of the $\pRpm(x,t)$ thus becomes as straightforward as for $\pLpm(x,t)$.
Once achieved, \eq{join} can then be used to construct a bipolar $\ppm(x,t)$
decomposition that satisfies {\em all three} conditions of
Sec.~\ref{additional}. On the other hand, from a purely practical standpoint,
little harm would result if one were to simply use $\pLpm(x,t)$
throughout $x$ and $t$. The reason is that, unlike
\eq{fdot4}, \eq{pdot} exhibits no coupling in {\em either}
$x$ asymptote. Thus, at both $t_0$ and $t_f$ for instance, $\pLpm(x,t)$
[and $\pRpm(x,t)$] evolve according to free particle propagation, which
introduces no nodes or interference. Condition 3. is therefore satisfied.
In practice, this is far more important than Condition 1., which---as
discussed above---is not satisfied for the $\pLpm^f$ decomposition
in the transmitted branch of $\psi^f(x)$.

\subsection{Multisurface generalization}
\label{multisurf}

Like the bipolar stationary state theory, the theory of bipolar wavepacket
dynamics can also be generalized for 1D multisurface applications. Let $f$
denote the number of electronic states considered. A diabatic-like
time-independent matrix \shro\ equation is presumed, of the form
\eb
    \tilde{H} \cdot \vec{\phi}^E = E \vec{\phi}^E,\label{multiSE}
\ee
where $\{\f_1,\f_2,...\f_f\}$ comprise
the vector components (associated with each of the $f$ diabatic
states) of the nuclear stationary state wavefunction, $\vec{\phi}^E$,
and
\eb
    \sof{\tilde{H}}_{i,j} = -\delta_{i,j} \of{{\hbar^2\over 2m}}
    \prtsq x + V_{i,j}(x) \label{multiH}
\ee
are the components of the $f\times f$ Hamiltonian operator
matrix, $\tilde{H}$, with $i\le f$ and $j\le f$ labeling diabatic
states.

The $V_{i,j}(x)=V_{j,i}(x)$ are the diabatic potential energy
curves, with the $i \ne j$ case denoting the coupling potentials. In
order to ensure that intersurface coupling vanishes in the asymptotic
limits (required to obtain asymptotic scattering waves with correct
boundary conditions),\cite{poirier06bohmIII,poirier07bohmIV,poirier07bohmalg}
we must have $V_{i\ne j}(x_L) = V_{i\ne j}(x_R) = 0$. However,
the asymptotic values for the diagonal potentials, $V_{i,i}(x)$, are
allowed to be completely arbitrary, and in particular, need not be
symmetric. Left and right asymptotic values are denoted $V_{iL} =
V_{i,i}(x_L)$ and $V_{iR} = V_{i,i}(x_R)$, respectively.

As per Sec.~\ref{asymmetric}, rather than work with generic
effective potentials $\Veff_i(x)$ that smoothly interpolate between
$V_{iL}$ and $V_{iR}$ values,\cite{poirier07bohmIV} we instead choose
constant effective potentials, $\Veff(x) = V^0$, with $V^0$
arbitrary for now. Note that in general, $V^0$ can be chosen to coincide
with at most one of the $V_{iL}$ and $V_{iR}$, so that we expect no more
than one of the 2$f$ component asymptotes to manifest perfect
asymptotic separation (Condition 1. from Sec.~\ref{additional}).

From \Ref{poirier07bohmIV}, the $\f_i = \f_{i+}+\f_{i-}$ components satisfy
\eb
     {\f_i}' = {i p \over  \hbar} \of{\f_{i+} - \f_{i-}}, \label{Fprime}
\ee
where $p = \sqrt{2m\of{E-V^0}}$. By combining \eq{multiSE} with
\eq{Fprime}, we obtain
\eb
    {\f_{i\pm}}' = \pm {i \over \hbar} \,p\, \f_{i\pm}
                   \pm {i \over \hbar} \of{{m \over p}} V^0\, \f_i
                   \mp {i \over \hbar} \of{{m \over p}} \sum_{j=1}^f \, V_{i,j}
                   \f_j, \label{fslashmulti}
\ee
the multisurface generalization of \eq{fslash}. Note that the
{\em same} constant $p$ is used for all components $i$, and in both
asymptotes $x_L$ and $x_R$.

By differentiating \eq{fslashmulti} with respect to $x$, and
otherwise applying the procedure described in Sec.~\ref{evolution},
we obtain time-evolution equations for the stationary state
components,
\eb
     {\partial \f_{i\pm} \over \partial t} =
                        - {i \over \hbar} \sum_{j=1}^f \sof{\tilde{H}}_{i,j} \f_{j\pm}
                         \mp {1 \over 2p} \sum_{j=1}^f V'_{i,j} \f_j, \label{fdot2multi}
\ee
the multisurface generalization of \eq{fdot2}. Finally, substitution of
the integral of \eq{Fprime} (with respect to $x$) into \eq{fdot2multi},
and integration over $E$ via an \eq{psidec}-type expansion yields the
following multisurface wavepacket time-evolution equations:
\ea{
     \lefteqn{{\partial \psi_{i\pm} \over \partial t} =}\hspace{3.0in}\nonumber\\
        - {i \over \hbar} \sof{\sum_{j=1}^f \sof{\tilde{H}}_{i,j} \psi_{j\pm}
              \pm\of{{1\over2}} \sum_{j=1}^f
                    V'_{i,j} \of{\Psi_{j+} - \Psi_{j-}}},
              \label{pdotmulti} & &}
where
\eb
              \Psi_{j\pm}  =  \int_{-\infty}^x
              \psi_{j\pm}(x')\,dx' \label{pintmulti}
\ee
Note that \eq{pdotmulti} satisfies the multisurface TDSE, in that
\eb
     {\partial \psi_{i} \over \partial t} = - {i \over \hbar} \bof{\sum_{j=1}^f
     \sof{\tilde{H}}_{i,j} \psi_{j}}.
\ee

For a given value of $V^0$, it is a straightforward matter to
propagate all of the $\psi_{i\pm}$ wavefunction components over
time, using \eq{pdotmulti} in conjunction with initial value conditions
discussed below. The resultant $\psi_{i\pm}(x,t)$
will in general satisfy Conditions 2. and 3. from
Sec.~\ref{additional}, but not Condition 1. Note that
the particular $\psi_i = \psi_{i+} + \psi_{i-}$ decompositions
obtained depend on the value of $V^0$, even though the time
evolution equations [\eq{pdotmulti}] are $V^0$-independent. As
described in Sec.~\ref{asymmetric}, the $V^0$ dependence manifests
in the initial conditions, $\psi_{i\pm}^0(x)$.

In addition to the conventions and conditions already adopted for
``well-behaved'' wavepacket dynamics in Sec.~\ref{additional},
let us further presume for the multisurface case that the initial
wavepacket is left-incident on surface $i=1$. Then,
$\psi_{(i>1)\pm}^0(x) = 0$, but $\psi_{1\pm}^0(x)$ depends
on $V^0$ via an \eq{pRinit}-type expansion
(with $\pRpm^0$ replaced with $\psi_{1\pm}^0$, $V_L$
replaced with $V_{1L}$, and $V_R$ replaced with $V^0$). One natural
choice for $V^0$ is $V^0=V_{1L}$ itself, leading to
$\psi_{1+}^0(x)=\psi_1^0(x)$ and $\psi_{1-}^0(x)= 0$.
This choice leads to perfect
asymptotic separation (i.e. Condition 1.) for $\psi_{1\pm}(x,t)$ in the
left asymptote, but generally not for $\psi_{1\pm}(x,t)$
in the right asymptote, nor for any of the $\psi_{(i>1)\pm}(x,t)$'s in
{\em either} asymptote.

We again reiterate that from a practical numerical perspective,
Condition 1. is not required---i.e. perfectly sensible results for
all $\psi_{i\pm}(x,t)$ may be obtained using the $V^0=V_{1L}$ choice
above, or any other reasonable $V^0$ value. On the other hand, if one
is determined to have perfect asymptotic separation for both left
and right asymptotes for all $i$, this can also be
achieved---via introduction of a dividing point $x_D$, and the
multisurface generalization of \eq{join}. In effect, this would
require that up to $2f$ separate calculations be performed,
corresponding to all of the distinct possibilities for $V^0=V_{iL}$
and $V^0=V_{iR}$. For each of these calculations, Condition 1. is
guaranteed for (at least) one $\psi_i$ component in one asymptote,
which is then singled out in that particular calculation---e.g.,
$\psi_{2R+}(x\!>\!x_D,t)$, from the $V^0 = V_{2R}$ calculation.
Finally, we note that for the asymptotically symmetric special case
considered in \Ref{poirier04bohmI}, where $V_{iL} = V_{iR} = 0$ for all $i$,
then the {\em single} choice $V^0=0$ leads to perfect separation in
both asymptotes for {\em all} components $\psi_{i}$.


\section{RESULTS}
\label{results}

We have applied the bipolar wavepacket time-evolution equations
derived in Sec.~\ref{theory} to a variety of model 1D applications.
The primary goal is to validate numerically that this approach
satisfies the three conditions of Sec.~\ref{additional}, especially
Condition~3. Consequently, little attention is paid here to
numerical efficiency, and only the simplest algorithms are employed,
using Eulerian fixed grids with uniform spacing in $x$ and $t$. No
trajectories or quantum potentials are computed; these will be
considered in later papers, that will actually solve the TDSE by
synthesizing quantum trajectories ``on the fly.''

In this paper, \eqs{pdot}{pdotmulti} are integrated over time using
the standard first-order forward Euler method, with fixed time step
size, $\Dlt$.\cite{press}  Eulerian fixed grids are used to
discretize the spatial coordinate $x$, with uniform spacing $\Dlt
x$, and left and right grid edges, $x_L$ and $x_R$, respectively.
Condition 2. from Sec.~\ref{additional} implies that Dirichlet
boundary conditions, $f(x_L) = f(x_R) = 0$, are employed, where
$f(x)$ represents any wavefunction component or its spatial
integral. The spatial derivatives implicit in the $\H$ contribution
to \eqs{pdot}{pdotmulti} are evaluted numerically using standard
symmetric (two-sided) second-order finite difference.\cite{press}
The spatial integrations are evaluated using closed Newton-Cotes
formulas---specifically, the two-point trapezoidal rule for the
second grid point from the left, and the three-point Simpson's rule
for all other interior grid points.\cite{press}

The initial wavepackets are all taken to be of the standard Gaussian form,
\eb
     \psi^0(x) = \of{2 \gamma \over \pi}^{1/4} \exp\sof{- \gamma (x-x_0)^2}
                \exp\of{{i p_0 x \over \hbar}}, \label{initial}
\ee
from which $\tilde{\psi}^0(p)$ can be determined analytically. In all
calculations, the parameters $\gamma$ and $p_0$ are chosen such that the
\eq{pcon} integral is negligibly small, i.e. comparable to the desired level
of numerical accuracy for the calculation, which is $10^{-6}$. Similarly,
all of the other numerical parameters, $\Dlt$, $\Dlt x$, $x_L$,
$x_R$, and $x_0$, are converged to the same level of accuracy. For the model
applications considered here, typical converged parameter values
in atomic units are as follows: $\Dlt \approx 0.1$; $\Dlt x \approx 0.08$;
$x_{L/R} = \mp 35$. Unless explicitly stated otherwise, the mass is taken
to be $m=2000$ a.u. Computer animations (.wmv file format) for all of the
wavepacket dynamics calculations presented in this paper are available as
EPAPS supplements,\cite{poirier07bohmVepaps}
and by direct request from the author.

\subsection{Eckart barrier system}
\label{eckart}

The canonical model scattering system for the asymptotically
symmetric special case is the Eckart barrier,\cite{eckart30,ahmed93}
defined via
\eb
     V(x)=V_0\,\text{sech}(\alpha x)^2, \label{eckartpot}
\ee
and specification of the parameters $V_0$, $\alpha$, and $m$.

\subsubsection{proton-like mass}
\label{eckartproton}

\begin{figure*}
\includegraphics[scale=0.85]{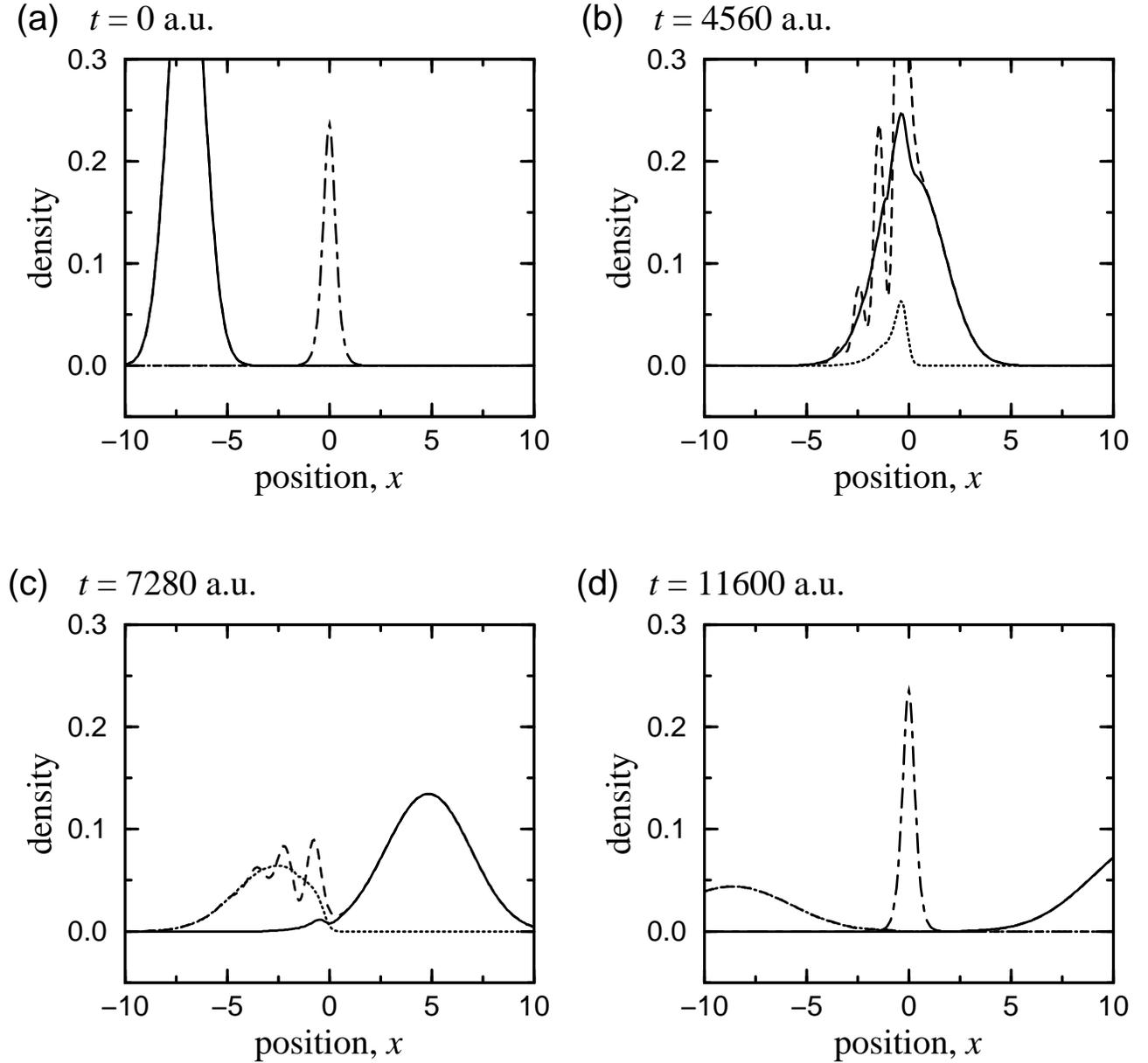}
        \caption{Wavepacket dynamics for the symmetric Eckart barrier system with $m=2000$ a.u. Each
                 plot represents a ``snapshot'' for a specific time, $t$, as listed (all units
                 are atomic units).
                 Various component wavepacket densities as a function of position are indicated
                 as follows: incident/transmitted, $\rho_+(x)= |\pp(x)|^2$, (solid);
                 reflected, $\rho_-(x)= |\pem(x)|^2$, (dotted); total, $\rho(x)= |\psi(x)|^2$,
                 (dashed). Initial and final densities, e.g. $\rho^0(x)$ and $\rho^f(x)$, are
                 presented in (a) and (d) respectively, in which the potential energy is
                 also represented (dot-dashed). In (b), the main peak of $\pp(x)$ has just
                 passed that of $\pem(x)$, though the spur (clearly visible) is forming.
                 The interference is mainly type I, and $\pem(x)$ is in stage 1. In (c),
                 $\pem(x)$ is in stage 2, and the interference is type II---caused by the
                 $\pp(x)$ spur, even though it has dissipated almost completely by this point.}
        \label{eckartprotonfig}
\end{figure*}

For the first Eckart application considered here, the parameter
values are chosen to be $V_0 = .0024$, $\alpha = 2.5$, and $m=2000$,
respectively, in atomic units. This is similar to what has been
called the ``Eckart A'' system in previous
papers.\cite{poirier06bohmIII,poirier07bohmIV,poirier07bohmalg}
In atomic units, the parameters
describing the initial Gaussian wavepacket of \eq{initial} are taken
to be $\gamma= 0.35$, $x_0 = -7.0$, and $p_0 = \,\,\,\sim\!\!3.28634$  a.u.
[Fig.~\ref{eckartprotonfig}(a)]. The $p_0$ value corresponds to an
incident kinetic energy of $.0027$ a.u., which is slightly above the
barrier energy $V_0$, so that substantial reflected and transmitted
branches of $\psi^f(x)$ are obtained.

\begin{figure}
\includegraphics[scale=0.6]{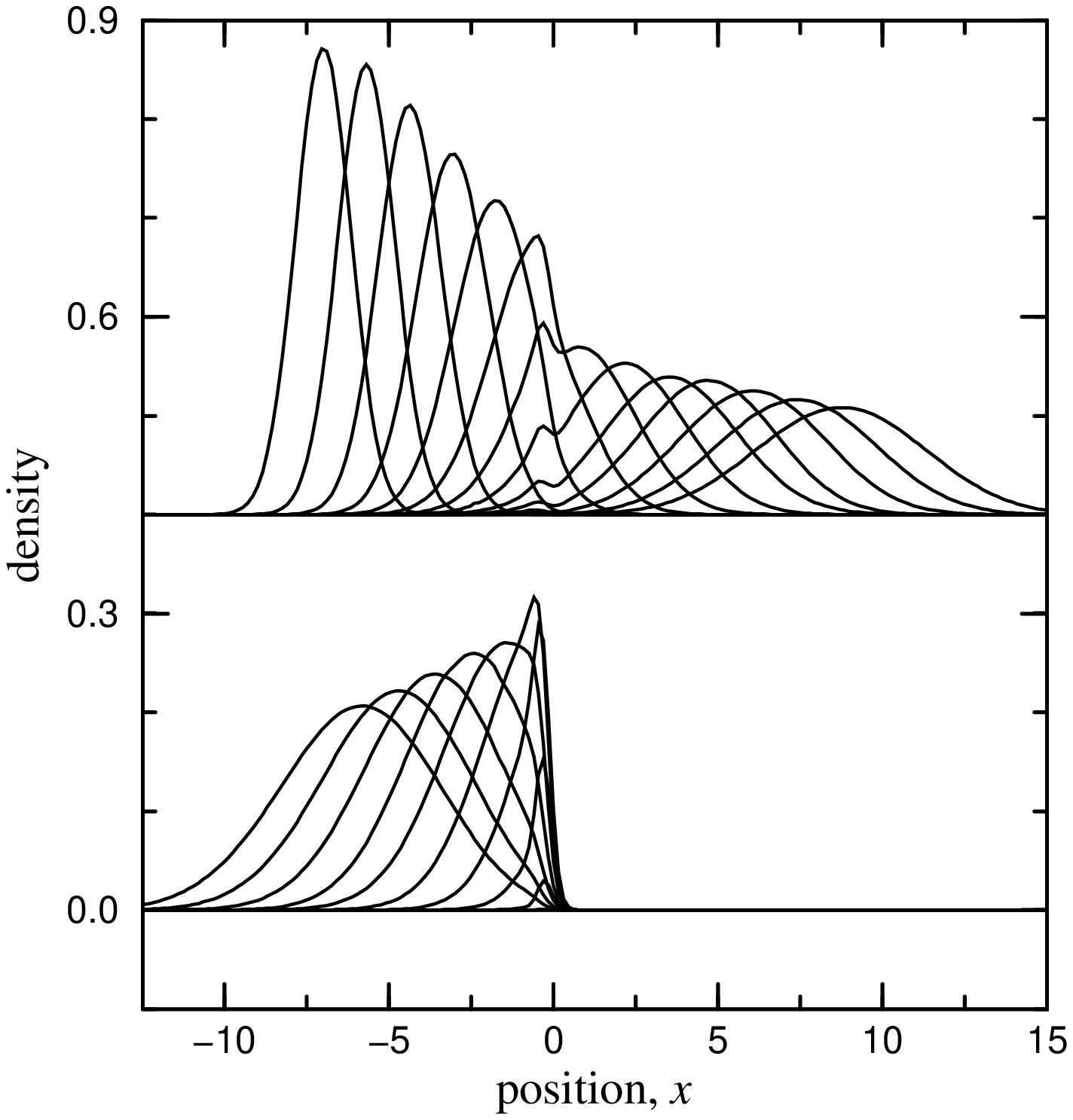}
        \caption{Component wavepacket densities, $\rho_{\pm}(x,t)$ as a function of position,
                 and for a variety of times, $t$, for the symmetric Eckart barrier system with
                 $m=2000$ a.u. The upper family of curves represent $\rho_{+}(x)$ at
                 different times, whereas the lower family of curves represent $\rho_{-}(x)$
                 (magnified by a factor of $4\times$). The motion of the former over time is
                 left-to-right, whereas that of the latter is right-to-left.
                 At intermediate $t$, a stationary $\rho_{+}(x)$ spur forms and then dissipates in
                 the interaction region, corresponding with the simultaneous ``birth'' and stationary
                 growth of $\rho_{-}(x,t)$. This is ``stage 1'' of the reflected wavepacket dynamics
                 (Sec.~\ref{eckartproton}), represented in the figure by the right-most four $\rho_{-}(x)$
                 curves. A sudden transition to stage 2 dynamics is then observed, in which $\rho_{-}(x)$
                 moves to the left.}
        \label{animationfig}
\end{figure}

The $\ppm(x,t)$ components are propagated using \eq{pdot}, and the
numerical methods described above, to a final time $\tmax =
(t_f-t_0) = 11600$ a.u., where reflected and transmitted branches
are found to be well-separated in left and right asymptotic regions,
respectively. Fig.~\ref{eckartprotonfig} shows the resultant wavepacket
dynamics for $\ppm(x,t)$ and $\psi(x,t)$ at four representative time
slices, including $t=t_0=0$ and $t=t_f=\tmax$. Although only four
time slices are presented, the bipolar decomposition has been
carefully inspected at all intermediate times
(Fig.~\ref{animationfig}), to ensure that Fig.~\ref{eckartprotonfig}
captures all of the relevant dynamics.

We find that Condition 1., Condition 2., and above all,
Condition 3. (from Sec.~\ref{additional}) are indeed well satisfied
 at all times. Both bipolar components $\ppm(x,t)$
are remarkably smooth, localized, non-oscillatory, and Gaussian-like
at all times---despite the fact that $\psi(x,t)$ itself displays a
substantial amount of interference at intermediate times
[dashed curves, Fig.~\ref{eckartprotonfig}(b) and (c)]. In addition to
interference, the proton-like mass is sufficiently small that the
wavepacket dynamics also manifest substantial {\em dispersion}, as
well. In fact---apart from a small ``spur'' that develops on
its left side (Sec.~\ref{spur}), and the fact that its integrated
probability decreases over time---the $\pp(x,t)$ evolution resembles
that of a {\em free particle Gaussian wavepacket}, simultaneously
translating and dispersing its way through the interaction region,
and in the process, smoothly transforming from the incident
wavepacket into the transmitted branch of the final wavepacket.

In contrast, the $\pem(x,t)$ wavepacket dynamics---though similarly
smooth and well-behaved---exhibit somewhat different behavior, due
to the different initial value conditions. In particular, the
$\pem(x,t)$ component is initially zero, but gradually comes
into being in the interaction region, as the $\pp(x,t)$
wavepacket passes through. The form of the coupling in \eq{pdot}
all but assures at least this much. In addition,
however---and quite unexpectedly---we also find that the $\pem(x,t)$
time-evolution undergoes two distinct stages. In the
first stage, $\pem(x,t)$ {\em stays in place} in the interaction region,
as it grows steadily in magnitude. Once the $\pem(x,t)$ integrated
probability has grown to roughly the final reflection probability value,
the second stage commences, in which $\pem(x,t)$ starts dispersing and
moving to the left, in roughly the same manner as for a free particle
Gaussian. This two-stage behavior---somewhat reminiscent of fruit ripening
on a tree, and then breaking free---is clearly evident in
Fig.~\ref{animationfig}, an ``animation plot''\cite{poirier07bohmIV} of the
time-dependent $\rho_{\pm}(x,t)$ densities.

Note that the transition from stage 1 to stage 2 occurs at a
time substantially {\em after} the
$\pp(x,t)$ peak has passed by that of $\pem(x,t)$. The resultant ``time
delay'' is a manifestation of the quantum
Goos-H\"anchen effect.\cite{hirschfelder74,jackson} The present
bipolar approach thus provides a means of measuring this effect {\em
directly}. The Goos-H\"anchen time-delay effect
appears to be closely related to the lack of combined flux
continuity discussed in Sec.~\ref{additional}. Note that for this
application, the total integrated probability,
$\int_{-\infty}^{+\infty} \of{\rho_+ + \rho_-}\, dx$, is fairly well
conserved over time, dipping down gradually from the
initial value of $1$ to a minimum value of $0.86$, and then
increasing to reach the final value of $1$ again by the end of the
propagation, as required.

\subsubsection{the $\pp$ ``spur,'' and its semiclassical interpretation}
\label{spur}
One very interesting and unexpected feature is the
small spur formed on the left side of $\pp(x,t)$ at intermediate
times. This spur remains behind the main $\pp$ peak, staying in
place above the growing $\pem$ wavepacket during stage 1., and then
moving ``backwards'' with $\pem$ during stage 2 (as is more evident
in calculations for parameters other than those used in
Sec.~\ref{eckartproton}). At a certain point in time, either before
or after the start of stage 2., the spur starts to diminish in size,
and eventually dissipates completely.  From
Fig.~\ref{eckartprotonfig}, the role of the spur is clear---i.e., to
bring about interference in the left, or ``reflected,'' part of
$\psi(x,t)$, at intermediate times.

It is well-known that the ``reflected'' part of $\psi(x,t)$ exhibits
nodes and interference to a far greater extent than the transmitted
part---indeed, one common strategy in traditional unipolar QTM
calculations is to ignore the reflected part altogether, after a
certain point in time.\cite{wyatt} To some extent, the
observed interference in $\psi(x,t)$ is due to the
the incident/transmitted and reflected contributions.
However, this cannot be the whole of the story, for the incident/transmitted
contribution [main $\pp(x,t)$ peak] often passes by,
long before the interference goes away completely. Indeed,
the so-called ``node healing'' process may continue even
{\em after} $\pem(x,t)$ has started to move away from the
interaction region. Eventually though, the nodes will be
healed completely---leading to a smooth final $\psi^f(x)$
reflected branch, and causing the $\pp(x,t)$ spur to vanish from existence.

It would thus appear natural to interpret the above as the
result of not two, but {\em three} separate contributions, as
considered briefly in Sec.~\ref{intro},  each of which is smooth and
well-behaved at all times.  In particular, the $\pem(x,t)$ component
is the reflected contribution, the main peak of $\pp(x,t)$ is the
incident/transmitted contribution, and the $\pp(x,t)$ spur---non-zero
only at intermediate times---is the third contribution. There are
thus two distinct types of interference: type I, between incident/transmitted
and reflected contributions, and type II, between spur and reflected
contributions. It is quite remarkable that the \eq{pdot} decomposition
should be generally capable of smoothly disentangling such varied and complex
interference patterns---a very delicate balance is evidently
required---yet this appears to be the case.

In Fig.~\ref{eckartprotonfig}(c), for instance, one can discern
interference in the $\rho(x)$ plot far to the left of the
interaction region, where $\rho_+(x)$ itself appears to be
vanishingly small. Yet this tiny $\pp(x)$ contribution is very
significant, for without it, $\rho(x)$ would equal $\rho_-(x)$,
which by visual inspection is clearly false. There is thus a
pronounced sensitivity in any \eq{psitot} bipolar
decomposition---owing ultimately to the square-root relation between
$\psi$ and $\rho$---which in practice, renders it exceedingly
difficult to completely disentangle interference effects.

\begin{figure*}
\includegraphics[scale=0.85]{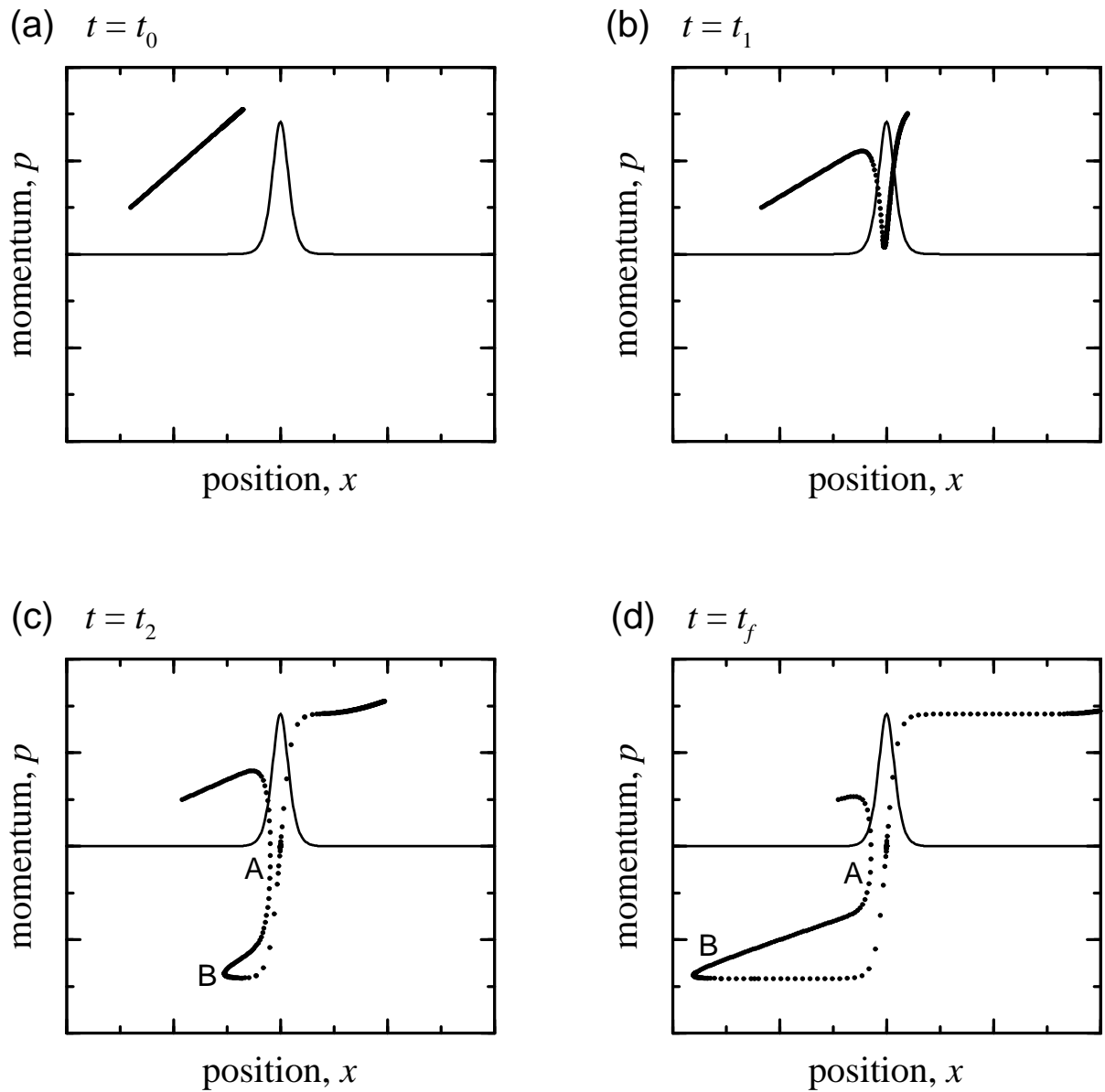}
        \caption{Semiclassical Lagrangian manifold (LM) dynamics for the symmetric Eckart barrier system.
                 Each plot represents a ``snapshot'' for a specific time, $t$. Thin solid curve
                 represents the Eckart potential. One-dimensional array of filled circles represents
                 discrete ensemble of classical trajectories, comprising the LM in phase space at each
                 point in time: (a) the initial $t=t_0$ LM is single-valued with respect to $x$; (b) by
                 $t=t_1$, the LM is still single-valued, but is about to form caustics; (c) at a later
                 time $t=t_2$, two caustics have formed, rendering the left side of the LM tripolar;
                 (d) caustic B moves to the left over time.}
        \label{classicalfig}
\end{figure*}

Returning to the idea of a {\em tripolar} decomposition, we comment
that further justification for this interpretation can be provided
using semiclassical arguments.\cite{poirier07bohmVunpub1}
In particular, consider an Eckart
barrier scattering problem for which the initial Gaussian wavepacket
is {\em spreading} (dispersing) at the initial time $t_0$ [unlike
\eq{initial}]. This stipulation is necessary in order that the
classical trajectories of the ensemble follow different orbits, so
that partial transmission and reflection can be achieved in a semiclassical
context. Figure~\ref{classicalfig} represents the time-evolution of such a
classical trajectory ensemble, or ``Lagrangian manifold''
(LM),\cite{tannor,poirier04bohmI,poirier06bohmII,poirier06bohmIII,keller60,maslov,huber88,littlejohn92}
in phase space.

The semiclassical LM, though initially single-valued,
develops caustics over time, rendering it multivalued at
later times. In particular, one caustic (labeled
``A'' in Fig.~\ref{classicalfig}) forms on the left of the barrier
peak, essentially sweeping through all of the individual trajectory
turning points, for those classical trajectories with insufficient
energy to clear the barrier. For a fairly narrow initial momentum
distribution, this caustic will not move very much in $x$-space,
over time. At a given time, those trajectories that have moved
through caustic A constitute the reflected branch of the
wavepacket, whereas those that have not comprise the
incident/transmitted wavepacket. In any event, caustic A is
responsible for type I interference.

Let us assume that the leading trajectories of the ensemble have
sufficient energy to clear the barrier---thus giving rise to the
transmitted branch of the LM at sufficiently large times. Since the
LM must be simply connected, there must be a {\em second} caustic to
the left of caustic A (labeled ``B''), which allows the LM to double
back through the interaction region and connect with the transmitted
branch, as indicated in the figure. Caustic B represents the leading
edge of the reflected wavepacket, and as such, continually moves to
the left.  Over time, the connecting thread between B and the
transmitted branch gets pulled apart by the separatrix ``like taffy,''
so that the integrated thread probability becomes vanishingly
small. The connecting thread itself therefore corresponds to the
$\pp$ spur, giving rise to type II interference, and a
{\em tripolar} LM representation in the ``reflected'' part of $\psi(x,t)$.
The present bipolar approach thus achieves classical
correspondence in the classical limit. In addition---as also observed
previously in the bipolar treatment of stationary
states\cite{poirier04bohmI,poirier06bohmII,poirier06bohmIII,poirier07bohmIV,poirier07bohmalg}---it
also leads to smooth, classical-like behavior far from the classical limit.


\subsubsection{electron mass}
\label{eckartelectron}

The general conclusions discussed in Secs.~\ref{eckartproton}
and~\ref{spur} have also been confirmed for a wide variety of
other parameter value choices for the Eckart barrier potential
[\eq{eckartpot}] and initial wavepacket [\eq{initial}],
including deep tunneling applications. In the interest of brevity,
we forego additional discussion of most of these additional calculations.
However, in light of the final comments in the preceding
Sec.~\ref{spur}, there is one particular case that merits
further attention---i.e., the quantum limit in which
$m\ra0$. To this end, we have performed bipolar wavepacket dynamics
calculations for the Eckart system using the {\em electron} mass $m=1$,
rather than the proton-like $m=2000$. Even in this extremely
non-classical regime, we find that the dynamical characterization
of $\ppm(x,t)$, as discussed in the preceding subsections still holds true.

\begin{figure*}
\includegraphics[scale=0.85]{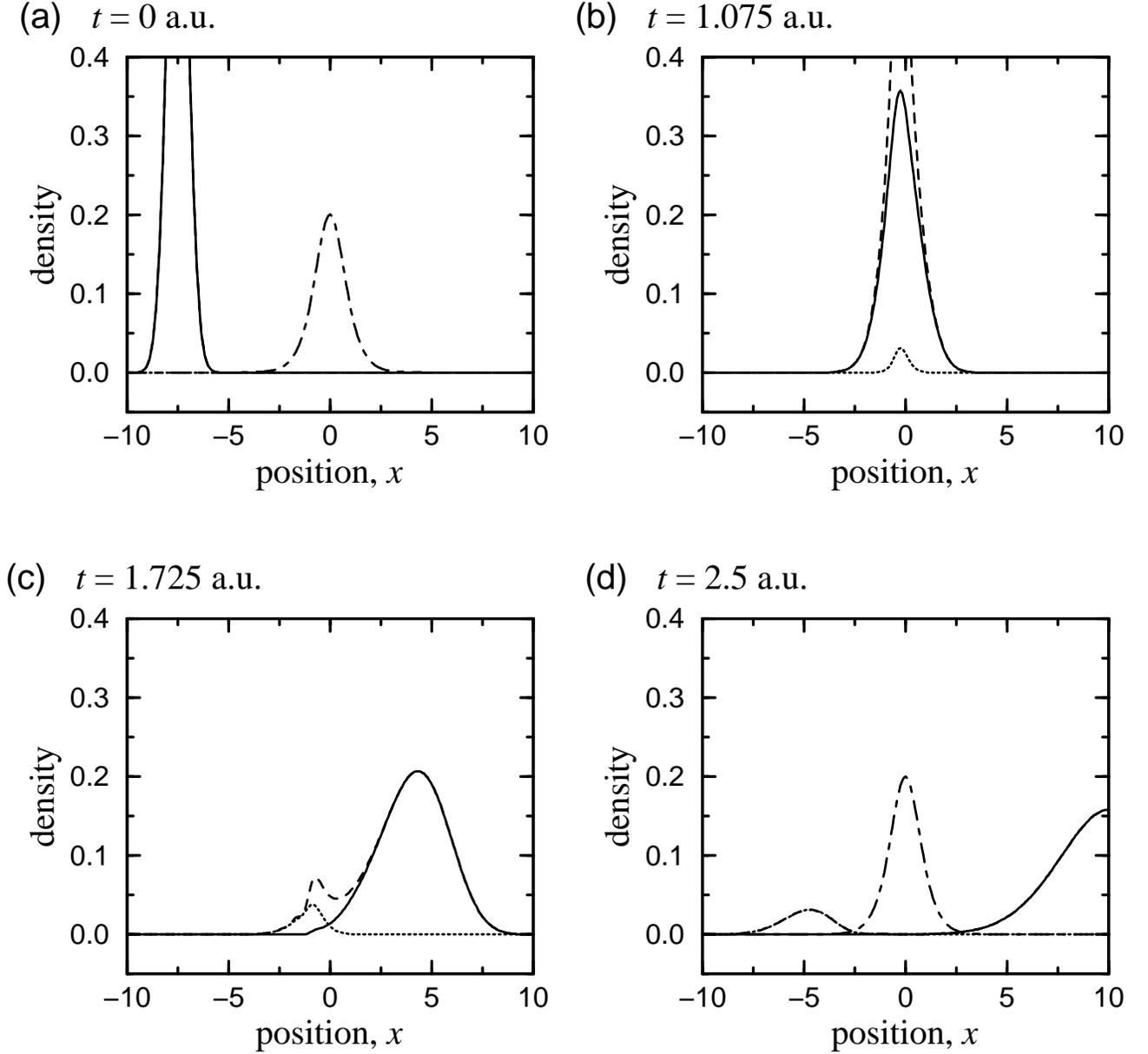}
        \caption{Wavepacket dynamics for the symmetric Eckart barrier system with $m=1$ a.u. Each
                 plot represents a ``snapshot'' for a specific time, $t$, as listed
                 (all units are atomic units). Various component wavepacket densities as a
                 function of position are indicated; see Fig.~\ref{eckartprotonfig} caption
                 for explanation. In (b), the $\pp(x)$ peak has just
                 passed that of $\pem(x)$, which is in stage 1. In (c), $\pem(x)$ has just
                 entered stage 2, implying a pronounced Goos-H\"anchen time delay [also
                 suggested by (d)]. Note the evident lack of interference, and of
                 a $\pp(x,t)$ spur.}
        \label{eckartelectronfig}
\end{figure*}

In atomic units, the particular parameter values used are as
follows: $V_0 = 20$; $\alpha = 1.0$; $m=1$; $\gamma = 1.0$; $x_0 =
-7.5$; $p_0 = \,\,\,\sim\!\!\!7.74597$.  The $p_0$ value corresponds to an
incident kinetic energy of 30 a.u., somewhat above the barrier peak,
but leading to substantial reflection and transmission.
Figure~\ref{eckartelectronfig} shows the resultant wavepacket
dynamics for $\ppm(x,t)$ and $\psi(x,t)$ at four representative time
slices, including $t=t_0=0$ and $t=t_f=\tmax=2.5$ a.u.  In comparison
with Sec.~\ref{eckartproton}, the dynamics is of course much faster,
and there is substantially less interference, as expected. Another
difference is that no $\pp(x,t)$ spur is evident, so that the interference
appears to be mainly of the type I variety.
In other respects, however, the situation is similar to the proton-like mass
case---in particular, all three conditions of Sec.~\ref{additional} are
clearly satisfied. This system also exhibits a substantial Goos-H\"anchen
time delay.

\subsection{Barrier ramp system}
\label{barrier}

The barrier ramp scattering system has
an asymmetric potential with a barrier. It serves as
a generic reaction profile for any direct chemical
reaction, and is thus an important benchmark system.
The potential functional form consists of an Eckart barrier
added to a hyperbolic tangent function, i.e.
\eb
     V(x)=V_0\,\text{sech}(\alpha x)^2 +
         \of{{V_R-V_L \over 2}} \sof{\tanh \of{\beta x}+1},
     \label{barrierpot}
\ee
such that the limiting values are $V(x_{L/R}) = V_{L/R}$,
as expected. In atomic units, the particular parameter values
used for the first barrier ramp calculation are as follows:
$V_0 = .0020$; $\alpha = \beta = 2.5$; $V_L=0$; $V_R=.0008$;
$m=2000$; $\gamma = 0.35$; $x_0 = -7.0$; $p_0 = 4$. These
parameters are chosen to correspond to those in Sec.~\ref{eckartproton},
except that a larger $p_0$ value is required, in order that \eq{pcon}
still be true with an upper limit of $\pmin$ rather than 0
(Sec.~\ref{asymmetric}). The $p_0$ value chosen corresponds to an
incident kinetic energy of $.004$ a.u., which is substantially
above the barrier peak of $\sim\!\!.0024$ a.u.

As per the discussion in Sec.~\ref{asymmetric}, two separate propagations
are performed, using \eq{pdot} for two different sets of initial conditions,
to a final time $\tmax = 9570$ a.u. The first propagation is for the
left $\pLpm(x,t)$ bipolar decomposition, corresponding to
$V^0 = V_L = 0$, for which $\psi_{L+}^0(x) = \psi^0(x)$
and $\psi_{L-}^0(x) = 0$. The second propagation is for the
right $\pRpm(x,t)$ decomposition,
for which the initial value condition is given by the \eq{pRinit} integration,
which is computed numerically using standard Fourier transform methods.
The two sets of solutions are then spliced together discontinuously via
\eq{join} at the dividing point, $x_D = 0$.

\begin{figure*}
\includegraphics[scale=0.85]{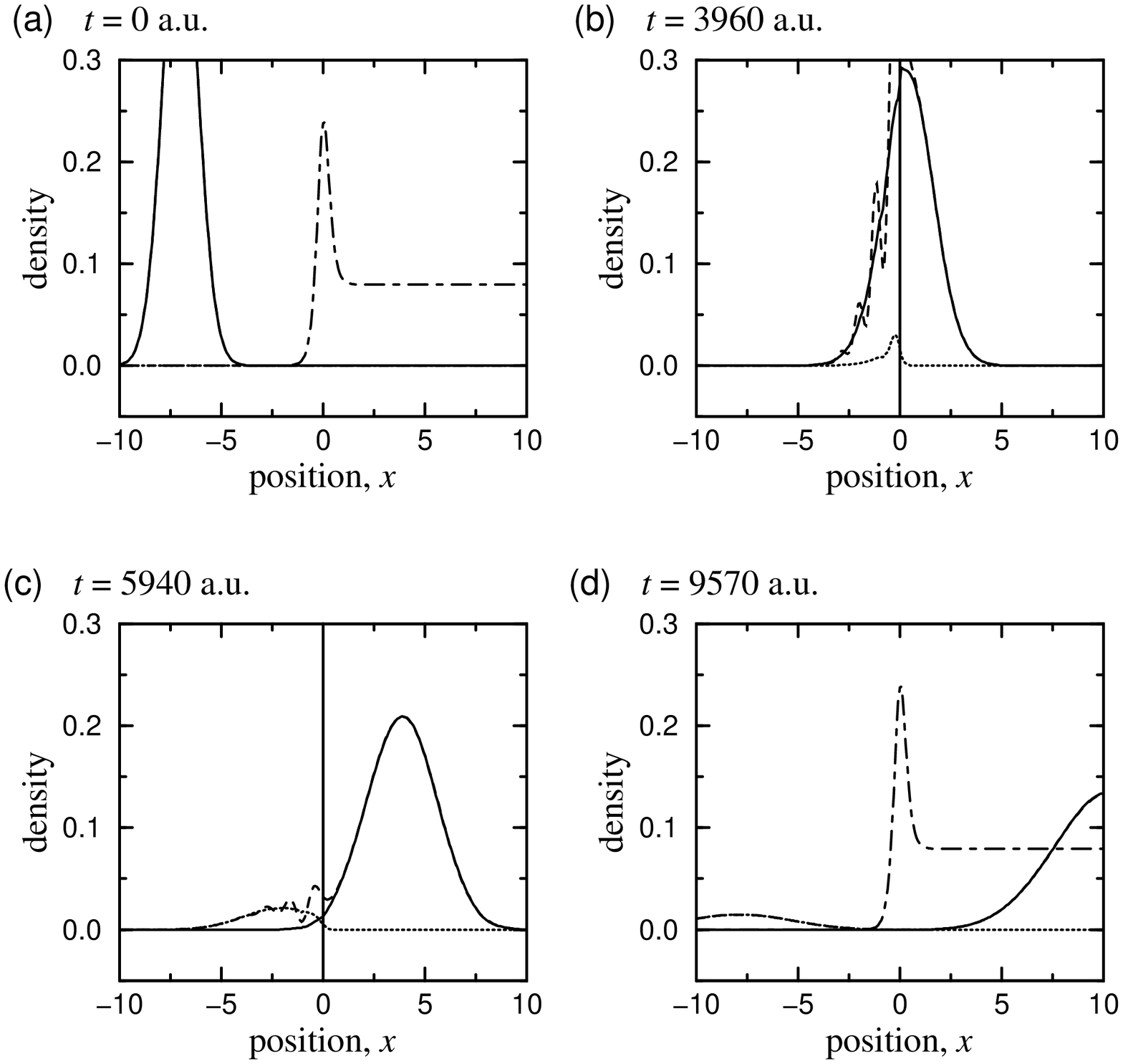}
        \caption{Wavepacket dynamics for the asymmetric barrier ramp system using ``spliced'' solutions,
                 $\ppm(x,t)$ [\eq{join}]. Each plot represents a ``snapshot'' for a specific time,
                 $t$, as listed (all units are atomic units). Various component wavepacket densities as a
                 function of position are indicated; see Fig.~\ref{eckartprotonfig} caption
                 for explanation. In (b) and (c), the vertical bar at $x=x_D=0$ indicates the
                 point where the $\pLpm(x,t)$ and $\pRpm(x,t)$ solutions are spliced together; the resultant
                 discontinuity is particular evident in the $\rho_+(x)$ plot of (c) (solid curve).}
        \label{barrierjoinfig}
\end{figure*}

Figure~\ref{barrierjoinfig} shows the
resultant wavepacket dynamics at four representative time slices.
The behavior is exactly as predicted in Sec.~\ref{asymmetric},
and otherwise comparable to that
of Sec.~\ref{eckartproton}---except that the reflected branch is smaller,
owing to the larger initial $p_0$ value. In particular, all three
conditions of Sec.~\ref{additional} are satisfied, and the bipolar
components $\ppm(x,t)$ are smooth and well-behaved---except of course
for the discontinuous join at $x=x_D$, most evident in the $\rho_{+}(x)$
plot of Fig.~\ref{barrierjoinfig}(b).  The $\ppm(x,t)$ are also
interference-free, though $\psi(x,t)$ itself exhibits substantial inferference
at intermediate times, as in Sec.~\ref{eckartproton}. Though difficult
to discern directly from the figure, $\psi_{L+}(x,t)$ forms a small spur,
which serves to heal the nodes in the reflected part of $\psi(x,t)$,
as observed for the Eckart barrier, and discussed in Sec.~\ref{spur}.

As an alternative to the above asymptotic joining procedure,
Sec.~\ref{asymmetric} also suggests simply working with the $\pLpm(x,t)$
solutions throughout all $x$ and $t$. The resultant wavepacket dynamics
are still anticipated to be smooth and well-behaved everywhere, though
Condition 1. will no longer be satisfied in the $(x>x_D,t_f)$ limit.
As an added benefit, moreover, it should be possible to work with initial
wavepackets with substantial $|\tilde{\psi}^0(p)|^2$ values in the
$0<p\le\pmin$ range, as this contribution is problematic only for the
$\pRpm(x,t)$ components, when $V_R>V_L$.

\begin{figure*}
\includegraphics[scale=0.85]{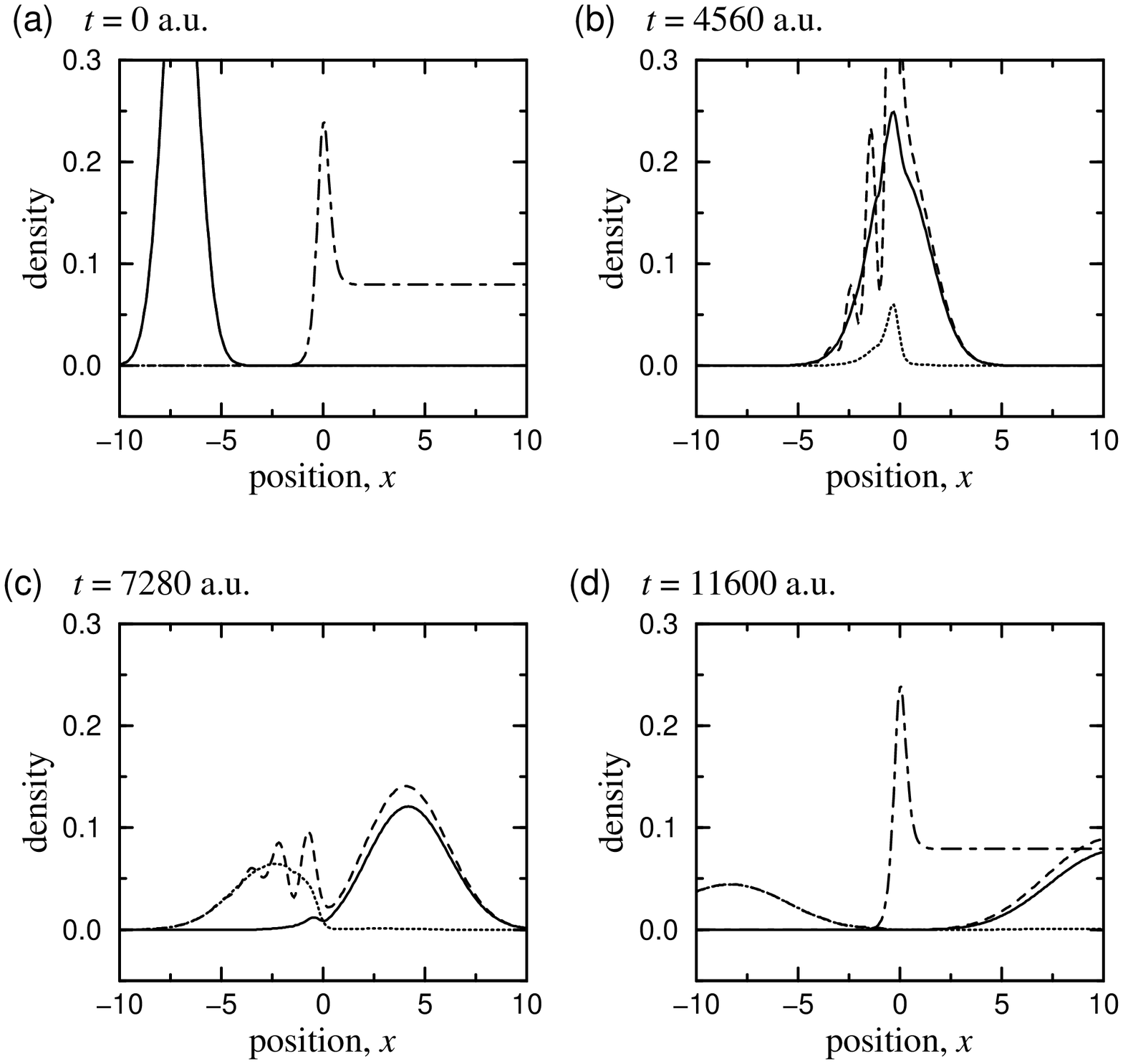}
        \caption{Wavepacket dynamics for the asymmetric barrier ramp system using $\pLpm(x,t)$ solutions
                 throughout $x$ and $t$. Each plot represents a ``snapshot'' for a specific time,
                 $t$, as listed (all units are atomic units). Various component wavepacket densities as a
                 function of position are indicated; see Fig.~\ref{eckartprotonfig} caption
                 for explanation. In comparison with Fig.~\ref{barrierjoinfig}, the component wavepacket
                 densities are now continuous everywhere, but do not satisfy perfect asymptotic separation
                 in the $(x>x_D,t_f)$ limit, as is evident in (c) and (d). Also, the reflection probability
                 is greater, due to the smaller value of $p_0$.}
        \label{barriernojoinfig}
\end{figure*}

To test this assumption, we have performed  $\pLpm(x,t)$ wavepacket
dynamics calculations for a second barrier ramp problem, using
initial and final wavepacket conditions {\em identical} to
Sec.~\ref{eckartproton}---i.e., $p_0 = \,\,\,\sim\!\!3.28634$  a.u.
and $\tmax = 11600$ a.u. All other parameters are as described
above. The results, presented in Fig.~\ref{barriernojoinfig}, are
very similar to those of Sec.~\ref{eckartproton}, except as
expected. In particular, both components are smooth and continuous
throughout, but there is a difference between $\rho(x>x_D)$ and
$\rho_+(x>x_D)$, which is clearly evident at large times---even
though $\rho_-(x>x_D)$ itself is barely visible.
[Fig.~\ref{barriernojoinfig}(d)]. This is a further manifestation of
the square-root sensitivity discussed in Sec.~\ref{spur}.

In addition to the above two barrier ramp problems, various other parameter
choices have been considered, including an electron-mass version analogous
to Sec.~\ref{eckartelectron}. In all cases, the resultant bipolar
decompositions have been found to be well-behaved at all times.


\subsection{Two-surface system}
\label{multiapp}

As our final test application, we consider a model multisurface
system with $f=2$ coupled electronic states. For this two-surface model,
both of the diagonal potential energy curves, as well as
the off-diagonal coupling potentials, are taken to be Eckart barriers, i.e.
\ea{
     V_{11}(x) & = & V_{22}(x) = V_0\,\text{sech}(\alpha x)^2,
                             \label{multipot} \\
     V_{12}(x) & = & V_{21}(x) = D_0\,\text{sech}(\alpha x)^2.}
In atomic units, the particular parameter values used are as follows:
$V_0 = .0024$; $D_0 = .00072$; $\alpha = 2.5$; $m=2000$;
$\gamma = 0.35$; $x_0 = -7.0$; $p_0 = \,\,\,\sim\!\!\!3.28634$.
These parameters are chosen to correspond to those used in
Sec.~\ref{eckartproton}, and also to yield substantial
final probability for all {\em four} components,
$\psi_{1\pm}^f(x)$ and $\psi_{2\pm}^f(x)$.

Note that the above model system conforms to the asymptotically
symmetric special case (end of Sec.~\ref{multisurf}).
Consequently, the choice $V^0=0$ leads to perfect asymptotic
separation for all four components (Condition~3). Since the
initial wavepacket is incident on surface $i=1$, the only
non-zero initial condition is for $\psi_{1+}^0(x)$, for which
\eq{initial} is used with parameter values given above. The four
components, $\psi_{1\pm}(x,t)$ and $\psi_{2\pm}(x,t)$, are then
propagated using \eq{pdotmulti}, to a final time $\tmax = 11600$ a.u.
Only a single propagation is required.

\begin{figure*}
\includegraphics[scale=0.85]{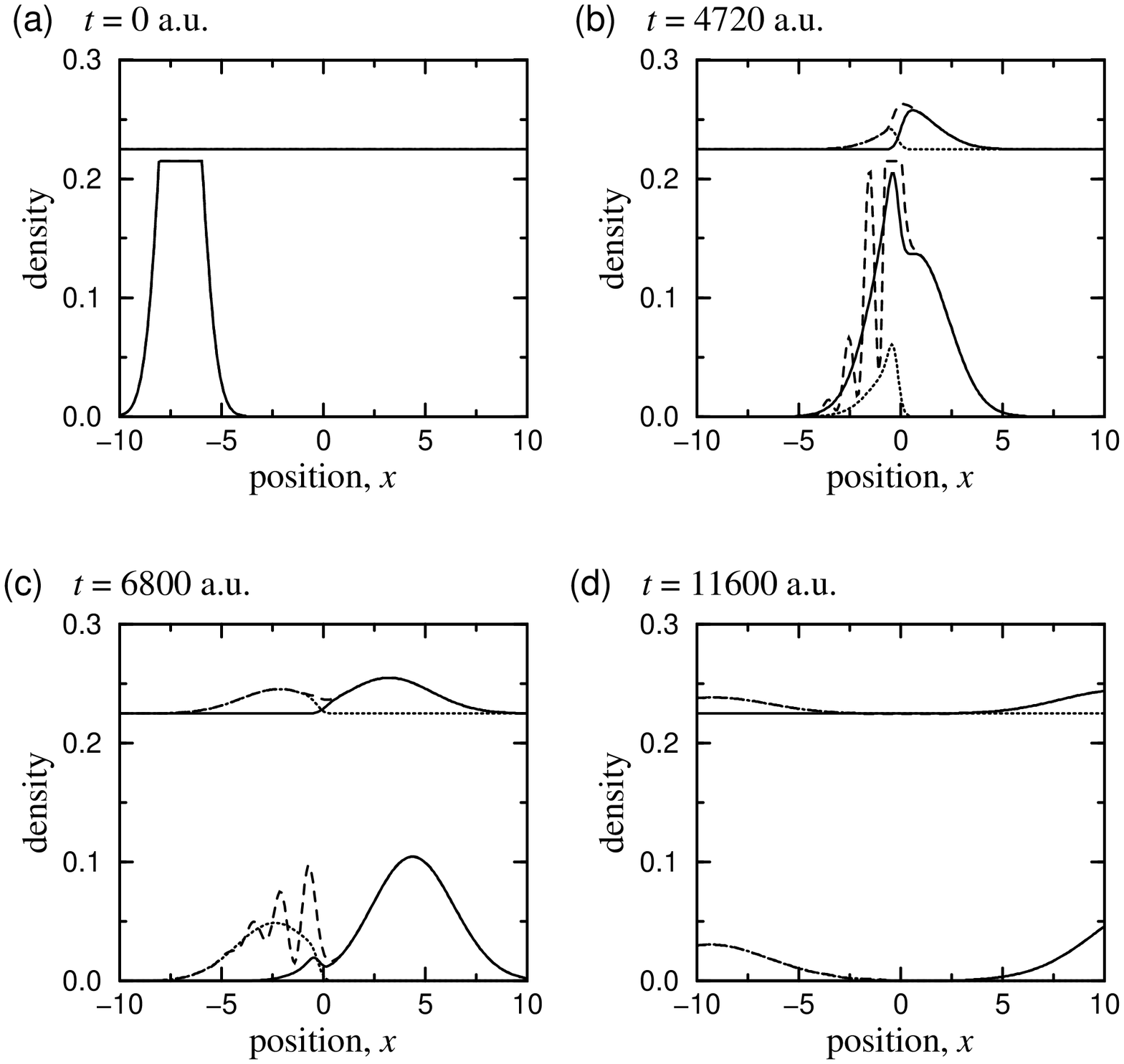}
        \caption{Wavepacket dynamics for the symmetric multisurface application of Sec.~\ref{multiapp}.
                 Each plot represents a ``snapshot'' for a specific time, $t$, as listed (all units
                 are atomic units). Various component wavepacket densities as a function of position
                 are indicated. Those for surface 1 are grouped together at the bottom of each plot
                 as follows: incident/transmitted, $\rho_{1+}(x)= |\psi_{1+}(x)|^2$, (solid);
                 reflected, $\rho_{1-}(x)= |\psi_{1-}(x)|^2$, (dotted); total, $\rho_1(x)= |\psi_1(x)|^2$,
                 (dashed). The corresponding surface 2 densities are grouped
                 at the top of each plot in similar fashion---e.g., transmitted,
                 $\rho_{2+}(x)= |\psi_{2+}(x)|^2$, via the upper solid curve, etc. Surface 2 exhibits little
                 or no interference, in part due to a lack of spurs. In contrast, the
                 $\psi_{1+}(x,t)$ spur is quite pronounced at intermediate times, i.e. in (b) and (c).}
        \label{multifig}
\end{figure*}

Figure~\ref{multifig} shows the resultant multisurface wavepacket dynamics
at four representative time slices. For visual clarity, the $\psi_{2\pm}(x,t)$
components are plotted above the $\psi_{1\pm}(x,t)$---though this is {\em not}
meant to imply that the $V_{22}$ potential is higher in energy. From the
figure, it is clear that each of the four components is smooth and
well-behaved at all $x$ and $t$, and otherwise acts as expected. As
the incident wave $\psi_{1+}(x,t)$ passes through the interaction region, all
{\em three} scattered components, $\psi_{1-}(x,t)$ and $\psi_{2\pm}(x,t)$,
come into being. All three of these components exhibit the two-stage process of
first growing in place, and then moving away from the interaction region
in their respective directions.

Note that, as in the earlier examples, $\psi_{1+}(x,t)$ develops a spur, which
provides type II interference (node healing) for the ``reflected'' part of
$\psi_1(x,t)$. The spur is particularly prominent for this system
[Fig.~\ref{multifig}(b) and (c)]. In contrast, neither
of the $\psi_{2\pm}(x,t)$ components develops a spur, so that
$\psi_2(x,t)$ has no type II interference. In fact, $\psi_{2}(x,t)$ does
not appear to exhibit any interference at all, even though the
$\psi_{2\pm}(x,t)$ components do overlap slightly at intermediate times,
which could in principle lead to type I interference. The reason is that
the $\psi_{2\pm}(x,t)$ are both growing in place, and therefore have the
same mean velocity of zero, whereas $\psi_{1+}(x,t)$ is moving relative
to $\psi_{1-}(x,t)$. This situation, though easily understood within
the present bipolar picture, would perhaps be difficult to justify in
purely semiclassical LM terms.

\section{SUMMARY AND CONCLUSIONS}
\label{conclusion}

This paper represents a turning point in the development of the
bipolar QTM methodology. All of the previous papers in the series have
focused exclusively on stationary states---whether bound or
scattering states, for either continuous or discontinuous
potentials. This was a necessary prerequisite for the present
paper, in addition to being useful in its own right---i.e., leading
to the development of extremely efficient and robust algorithms for
1D stationary scattering applications (\Ref{poirier07bohmIV} and
\Ref{poirier07bohmalg}). These algorithms have been succesfully
applied to challenging deep tunneling applications, and also to
1D reaction path Hamiltonian approximations for real chemical reactions
such as Cl$^-$ + CH$_3$Cl $\ra$ ClCH$_3$ +  Cl$^-$. On
the other hand, the direct calculation of the stationary $\phi^E$
states, as functions of position, will never be feasible for very
large systems, owing to the fact that such states are delocalized
over a configuration space of many dimensions. If such systems are
to succumb to exact quantum dynamical treatment in a position-space
representation, it must be via some non-stationary, localized
wavepacket approach. Moreover, the dynamical equations used for the
numerical wavepacket propagation must be free of any explicit
dependence on the $\phi^E$ states, or even $E$ itself.

This we have achieved here in
\eqs{pdot}{pdotmulti}---through a sequence of
manipulations of the original time-evolution equations for
stationary scattering states [\eq{fdot}], and the introduction of
reasonable restrictions on the allowed wavepacket dynamics
(Sec.~\ref{additional}). The intermediate result, \eq{fdot2},
though not used directly, depends on the JWKB quantity
(Sec.~\ref{evolution})---thereby affording a
theoretical connection with semiclassical mechanics and the
classical limit. The final \eq{pdot} is an integro-differential
equation that depends on the
spatial {\em integral} of $\ppm$, rather than the integral of
$|\ppm|^2$, or $\ppm$ itself. This is quite unusual in the context of
quantum dynamics---though commonplace, for instance, in
soliton dynamics.\cite{dauxois,poirier07bohmVprivate}
In any case, it is precisely this spatial integration that
removes all explicit dependence of \eq{fdot3} on $E$ and
$p$---the crucial requirement in the derivation of \eq{pdot}.

The derivation is more involved than originally anticipated, owing
to the fact that \eq{fdot} is presented in a form that is unsuitable
for a wavepacket generalization---even though ultimately, \eq{pdot}
is derived from \eq{fdot}, and is thus equivalent to it
(at least at asymptotically large times). Nevertheless, there are essential
differences between the two equations, stemming from the fact that the coupling
contribution in \eq{fdot} is not the same as in \eq{fdot3}. One important
difference is that \eq{pdot} is independent of the constant potential value,
$V_0$, used to define the stationary state trajectories (though the $V_0$ value
does affect the initial value conditions). Also, the TDSE is satisfied at
all times (not just asymptotically large times). Finally, unlike
\eq{fdot}, \eq{pdot} does not conserve total integrated probability
[\eq{fluxrel}].

Another important new development is that the bipolar wavepacket
methodology readily lends itself to a traditional
Bohmian mechanics interpretation---e.g., quantum trajectories
defined via $S_\pm' = p_\pm$ [\eq{psipm}], with dynamics governed by the
quantum potential.  The behavior of the bipolar quantum
trajectories ensuing from \eq{pdot} will be explored in a future
publication. In practical terms, however, the trajectories are
essentially guaranteed to be well-behaved, provided that all three of
the conditions for well-behaved bipolar wavepacket components,
as described in Sec.~\ref{additional}, are satisfied for all
$x$ and $t$. Condition 3., in particular, is required to circumvent
the node problem plaguing the synthetic QTM approach,
and is therefore essential for application to large systems, for which
the Eulerian fixed-grid algorithm used here (however efficiently
implemented) would be unfeasible.

To satisfy all three conditions of Sec.~\ref{additional} for a wide
variety of applications is decidedly nontrivial, and quite a lot to
ask of any set of dynamical equations---especially with regard to
the interference Condition 3., owing to the square-root sensitivity
discussed in Sec.~\ref{spur}. Indeed, it may be the case that {\em
no} set of evolution equations other than \eq{pdot} would be capable
of achieving such a separation for general applications. In
fact, a great many candidates were explored and rejected, for
failing to satisfy Condition. 3 (and in many cases, Conditions 1.
and 2. as well).\cite{poirier07bohmVunpub2}
The examples of Sec.~\ref{results} (and others)
nevertheless indicate that the present bipolar wavepacket approach
appears capable of achieving this---in both the quantum and
classical limits, and even when there is a complicated interplay of
at least two different types of interference. These examples also
clearly demonstrate that virtually all quantum effects that play a
role in wavepacket dynamics---i.e. dispersion, tunneling, and
interference---can be easily incorporated. Note that for real
molecular systems with atomic nuclei heavier than
hydrogen, interference effects can be even more pronounced than for
any of the examples considered in this paper; such cases will be
considered in future publications.

The present work thus serves to demonstrate that a synthetic bipolar
wavepacket QTM approach based on \eq{pdot} would be widely applicable and
numerically feasible---using the quantum potential to handle those
quantum effects that it does best, i.e. dispersion and
tunneling, and intercomponent coupling to treat interference. The
numerical algorithm would employ well-established unstructured grid
techniques, such as local least-squares
fitting,\cite{wyatt,wyatt01b,wyatt99} that have already been applied
successfully to standard unipolar wavepacket QTM applications {\em
without} interference. In the unipolar context, the chief numerical
requirement is the calculation of spatial derivatives needed to
compute the quantum potential or quantum force. The only new numerical
requirement for the bipolar treatment is that of spatial {\em
integration}, to evaluate $\Ppm(x,t)$ at every time step. This is not
anticipated to pose severe numerical difficulties, even for multidimensional
applications, as the generalization of \eq{pint} is a {\em line}
integral (1D), rather than a volume integral over all degrees of freedom.

As a practical matter, it is important that the present approach can be
generalized for arbitrary asymmetric potentials, and even multisurface
applications, as discussed in Secs.~\ref{asymmetric} and~\ref{multisurf},
respectively.  Intriguingly the time-evolution equations for the former
are unmodified from the symmetric case, although the initial and final value
conditions do depend on the asymptotic potential values. From a
formal perspective, the fact that multiple calculations must
be performed in the asymmetric case---and then ``glued'' together
discontinuously at the dividing point, $x_D$---is perhaps less
appealing than the continuous $\Veff$ approach developed for stationary
state applications in \Ref{poirier07bohmalg}. On the other hand, the
discontinuous approach is also taken by other standard ``dividing surface''
reactive scattering methods, which essentially posit two completely
different asymptotic Hamiltonians, one for reactants, and one for
products.\cite{seideman92a,rom92} In any event, a generalization of the bipolar
wavepacket theory using step effective potentials
$\Veff(x)= V_L + (V_R-V_L) \Theta(x-x_D)$
directly (\Ref{poirier06bohmIII}), will be explored in a future paper. This will
allow direct propagation across the discontinuity, which in turn, implies
that only a single calculation need be performed for asymptotically
asymmetric systems, while still satisfying perfect asymptotic separation.
Another issue that may be considered in future is the generalization for
initial wavepackets that do {\em not} satisfy \eq{pcon}---i.e., that include
an initially outgoing contribution. The nominal difficulty is that such
wavepackets lead to delocalized $\Psi$ functions, although this issue may be
resolved by explicit consideration of the {\em right}-incident stationary
solutions.

In addition to the various ``horizontal'' developments, to be considered
in future publications as described above, the next step
in the ``vertical'' or methodological direction---and the
subject of the final paper in this series---remains the development of
a {\em multidimensional} generalization. In a sense,
\eq{pdotmulti} already provides one avenue for
multidimensional application---in that the multiple surfaces may be
regarded as parametrized, discrete quantum states for all of the
``perpendicular'' degrees of freedom. This approach could be used, for
instance, to treat rotational degrees of freedom in the context of a
partial wave expansion. In paper~VI though, we shall
address wavepacket dynamics directly on the full-dimensional
configuration space. The theoretical development required will be
seen to be a remarkably straightforward generalization of that
presented here. In particular, only {\em two} wavefunction components
are still required---regardless of system dimensionality---provided
there is a single reaction path. Moreover, the formalism can
accommodate standard Jacobi-type coordinate representations
with arbitrary curvilinear reaction paths. Paper~VI will present
results for several multidimensional applications including
collinear H+H$_2$---the first such exact quantum dynamics
calculations ever performed using a QTM. Moreover, the resultant
multidimensional $\ppm$ decomposition will be found to satisfy
the three all-important conditions of Sec.~\ref{additional}.

As a final observation, we note that the bipolar wavepacket
decomposition as presented in this paper is by no means restricted to
the Bohmian mechanics context only. More generally, it ought to be
regarded as a scattering formalism in its own right, which could in
principle impact favorably on any of a number of existing
computational methodologies. Even straightforward Eulerian
fixed-grid propagations, for instance, might benefit from the
greatly increased grid spacing and time step sizes associated with
the component $\ppm(x,t)$'s---which are much smoother in general than
$\psi(x,t)$ itself, particularly in the classical limit.
Another idea might be to exploit the
Gaussian-like properties of the component $\ppm(x,t)$'s to develop
an approximate Gaussian evolution scheme, \`a la
Heller.\cite{tannor,heller81} Perhaps as few as three such ``growing
Gaussians'' would be required---one for the incident/transmitted
wavepacket, one for the reflected wavepacket, and a third for the
spur.

\begin{acknowledgments}

This work was supported by a grant from The Welch Foundation (D-1523),
and by a Small Grant for Exploratory Research from the National Science
Foundation (CHE-0741321).
The author wishes to express gratitude to Yair Goldfarb,
Salvador Miret-Art{\'e}s, Lucas Pettey, Angel Sanz,
David Tannor, and Robert Wyatt, for many interesting discussions.
The author is particularly indebted to George Hinds and Jeremy
Schiff for discussions pertaining to the possible relationship
between quantum dynamics and solitons. Lucas Pettey, and especially
Corey Trahan are also acknowledged, for  Herculean efforts to
implement a great many of the earlier, unsuccessful ideas.
Jason McAfee is also acknowledged for his aid in converting this manuscript to an electronic format suitable for the arXiv preprint server.

\end{acknowledgments}

%
%

%
%
%



%
%

\end{document}